\documentclass{French}
\usepackage{latexsym}
\usepackage[french]{babel}
\usepackage{psfrag}
\usepackage{amsmath}
\usepackage{amssymb}

\begin{document}
\subh{Two competing improvements of PID controllers$\Huge{:}$ A comparison 
}
\author{Michel Fliess\ts{1,3} et Cédric Join\ts{2,3,4}}
\aff{\removelastskip \ts{1}LIX (CNRS, UMR 7161), \'Ecole polytechnique, 91128 Palaiseau, France, Michel.Fliess@polytechnique.edu\\
\ts{2}CRAN (CNRS, UMR 7039), Université de Lorraine, BP 239, 54506 Vand{\oe}uvre-lès-Nancy, France, Cedric.Join@univ-lorraine.fr\\
\ts{3}AL.I.E.N. (ALgèbre pour Identification \& Estimation Numériques), 7 rue Maurice Barrès, 54330 Vézelise, France, \{michel.fliess, cedric.join\}@alien-sas.com \\
\ts{4}Projet Non-A, INRIA Lille -- Nord-Europe, France}

\resu{Aujourd'hui, \og{commande sans modèle}\fg, ou \og{MFC}\fg, et \og{commande par rejet actif de perturbations}\fg, ou \og{ADRC}\fg, sont les approches les plus en vue pour préserver les avantages des PID, si populaires dans l'industrie, tout en atténuant leurs carences. Après un bref rappel sur MFC et ADRC, plusieurs exemples démontrent la supériorité du sans-modèle car permettant d'embrasser une classe beaucoup plus vaste de systèmes.} 
\abstract{In today's literature \emph{Model-Free Control}, or \emph{MFC}, and \emph{Active Disturbance Rejection Control}, or \emph{ADRC}, are the most prominent approaches in order to keep the benefits of PID controllers, that are so popular in the industrial world, and in the same time for attenuating their severe shortcomings. After a brief review of MFC and ADRC, several examples show the superiority of MFC, which permits to tackle most easily a much wider class of systems.}
\moi{\textbf{\textcolor{AbsBlue}{MOTS-CL\'{E}S.}}\hphantom{--} PID, commande sans modèle, commande par rejet actif des perturbations, commande par platitude}
\kwd{PID, model-free control, MFC, active disturbance rejection control, ADRC, flatness-based control}

\chapter{Deux améliorations concurrentes des PID}
\begin{flushright}
Ce que l'esprit voit le c{\oe}ur le ressent. \\
Malraux (\textit{La Condition humaine}, Paris: Gallimard, 1933)
\end{flushright}

\section{Introduction \vspace*{-5truemm}
}
On conna\^{\i}t la domination, écrasante dans l'industrie, des correcteurs de type \og{proportionnel-intégral-dérivé}\fg, ou \og{PID}\fg \ (voir, par exemple, \cite{astrom},  \cite{murray}, \\ \cite{od}). \`A côté d'avantages aussi considérables que reconnus, singulièrement la simplicité conceptuelle et l'inutilité d'une modélisation mathématique, des manques flagrants conduisant trop souvent à une mise en {\oe}uvre pénible et à des performances médiocres. La recherche universitaire, si active en automatique, dite parfois \og{moderne}\fg \ depuis plus de cinquante ans, y trouve en bonne part sa motivation. Elle suppose fréquemment une description par équations différentielles ou aux différences. Ses contributions, théoriques et pratiques, sont incontestables (voir, par exemple, \cite{murray}, \\ \cite{kumar}). Afin de ne pas trop alourdir cette publication, contentons-nous, ici, de citer deux apports, aux applications aussi nombreuses que diverses: le filtre de Kalman \cite{kalman1,kalman2} et la platitude \cite{flmr1,flmr2}.

Durant toutes ces années, on a tenté à maintes reprises, avec, semble-t-il, un succès mitigé jusqu'à récemment, de garder les avantages des PID tout en gommant leurs défauts. Deux voies, plutôt nouvelles, prévalent à l'heure actuelle:
\begin{enumerate}
\item la \og{commande sans modèle}\fg \ \cite{ijc13}, ou \emph{model-free control}, désignée ici par \emph{MFC}, son acronyme anglais;
\item la \og{comande par rejet actif de perturbations}\fg, ou \emph{active disturbance rejection control}, désignée aussi par son acronyme anglais, \emph{ADRC}. Cette méthode, issue avant tout des travaux de \cite{han} en Chine, a été développée par divers auteurs. Voir, par exemple,
\begin{itemize}
\item les articles signés par \cite{aguil}, \cite{feng}, \cite{gao}\footnote{Certains auteurs, comme \cite{gao}, ont voulu faire remonter quelques-uns des principes de l'ADRC plus haut dans le temps, à \cite{poncelet} en particulier.}, \\ \cite{pde}, \cite{huang}, \cite{inoue}, \cite{qi}, \\ \cite{madon2}, \cite{polo}, \cite{tav}, \\ \cite{vincent}, \cite{wu}, \cite{yao},  \cite{zhang}, \\ \cite{zheng0});
\item les livres de \cite{guo} et \cite{sira3}.
\end{itemize}
\end{enumerate}
MFC et ADRC ont permis bien des applications, parfois spectaculaires. Les bibliographies des références plus haut sur l'ADRC en fournissent une liste assez complète. Pour le MFC, renvoyons à la bibliographie de \cite{ijc13}, à \cite{alinea,bldg} et à leurs références.

Des adeptes de l'ADRC ont publié des critiques, aussi vives qu'infondées, contre le MFC (voir, par exemple, \cite{madon1}). L'ignorance totale chez d'autres, comme \cite{guo}, des techniques du MFC aboutit à un exposé partial. Certains, comme \cite{cortes}, effacent l'écart béant les séparant.  Cet article use d'un droit de réponse légitime. Il vise à rétablir les faits, en exploitant, après avoir résumé les deux approches, plusieurs exemples académiques\footnote{Aux lecteurs de comparer les applications concrètes grâce aux références bibliographiques. Une telle évaluation est, dans le cadre de cet article, impossible.}. Puisse une discussion ouverte et fructueuse s'ensuivre\footnote{L'importance pour l'ADRC de la contribution chinoise est décisive. On doit souligner d'autant plus la grande qualité de maints travaux en Chine sur le MFC (voir, par exemple, \cite{ticher,wang17,wang16,delayzhang}).}.

Le paragraphe 2 résume MFC et ADRC. On s'y inspire pour l'ADRC du nouveau livre \\ \cite{sira3}, plus clair. Le suivant exhibe des simulations numériques,  qui vont du linéaire au non-linéaire en terminant par une équation aux dérivées partielles. Ils tendent à prouver la supériorité du sans-modèle. La conclusion évoque aussi: 
\begin{enumerate}
\item l'influence sur la compréhension de questions naturelles; 
\item les retards; 
\item les équations aux dérivées partielles. 
\end{enumerate}

\section{Les deux approches}\label{rap}
\subsection{MFC}\label{mfc}

\subsubsection{\itshape Emploi d'un théorème d'approximation}
On se restreint, afin de simplifier l'écriture, à un système \emph{SISO}\footnote{Acronyme de \emph{Single-Input Single-Output}.}, c'est-à-dire avec une seule commande $u$ et une seule sortie $y$. On suppose la \og{fonctionnelle}\fg \ correspondant à $y$, c'est-à-dire la fonction de la fonction $u$, \emph{causale}, ou
\emph{non-anticipative}: 
\begin{equation}\label{functional}
\forall t > 0, \quad y(t) = \mathcal{F}\left( u(\tau) ~ | ~ 0 \leqslant \tau \leqslant t \right)
\end{equation}
$\mathcal{F}$ dépend
\begin{itemize}
\item du présent et du passé, mais pas du futur;
\item de perturbations variées;
\item de conditions initiales en $t = 0$.
\end{itemize}

\textbf{Exemple}.
Populaires en automatique (voir, par exemple, \cite{lamnabhi,rugh}) et dans divers domaines des sciences appliquées, les \og{séries de Volterra}\fg \  sont des fonctionnelles du type 
\begin{align*}\label{vs}
y(t) = & h_0(t) + \int_0^t h_1(t, \tau) u(\tau) d\tau + \\ &
\int_0^t \int_0^t h_2(t, \tau_2, \tau_1) u(\tau_2) u(\tau_1) d\tau_2
d\tau_1 + \dots \\
& \int_0^t \dots \int_0^t h_\nu(t, \tau_\nu, \dots, \tau_1)
u(\tau_\nu) \dots u(\tau_1) d\tau_\nu \dots d\tau_1  \\ & + \dots
\end{align*}
Elles apparaissent, notamment, comme solutions d'équations différentielles ordinaires assez générales (voir, par exemple, \cite{lamnabhi}, \cite{rugh} et \cite{volt}).

On introduit
\begin{itemize}
\item un compact $\mathcal{I} \subset [ 0, + \infty [$;
\item un compact $\mathcal{C} \subset C^0 (\mathcal{I})$, où $C^0 (\mathcal{I})$ est l'ensemble des fonctions continues $\mathcal{I} \rightarrow \mathbb{R}$, muni de la topologie de la convergence uniforme.
\end{itemize}
Soit $\mathfrak{S}$ la $\mathbb{R}$-algèbre de Banach des fonctionnelles \eqref{functional}, causales et continues, $\mathcal{I} \times \mathcal{C} \rightarrow \mathbb{R}$.
D'après le théorème de Stone-Weierstra{\ss} (voir, par exemple, \cite{choquet,rudin}), toute sous-algèbre, contenant une constante non nulle et séparant les éléments de $\mathcal{I} \times \mathcal{C}$, est dense dans $\mathfrak{S}$.
Soit $\mathfrak{A} \subset \mathfrak{S}$ l'ensemble des fonctionnelles satisfaisant des équations différentielles algébriques du type
\begin{equation}\label{eq}
E(y, \dot{y}, \dots, y^{(a)}, u, \dot{u}, \dots, u^{(b)}) = 0
\end{equation}
où $E$ est un polynôme à coefficients réels. Selon \cite{ijc13}, $\mathfrak{A}$ est dense dans $\mathfrak{S}$. 
Il est ainsi loisible de supposer que le système considéré est \og{bien}\fg \ approché  par un système \eqref{eq}. Soit un entier $\nu$, $1 \leqslant \nu \leqslant a$, tel que
$$
\frac{\partial E}{\partial y^{(\nu)}} \not\equiv 0
$$
D'où, localement, d'après le théorème des fonctions implicites,
$$
y^{(\nu)} = \mathcal{E}(y, \dot{y}, \dots, y^{(\nu - 1)}, y^{(\nu +
1)}, \dots, y^{(a)}, u, \dot{u}, \dots, u^{(b)})
$$
Il en découle le système \og{ulta-local}\fg
\begin{equation}
\boxed{y^{(\nu)} = F + \alpha u} \label{ultralocal}
\end{equation}
Souvent, en pratique, on sélectionne $\nu = 1$. Voir \cite{ijc13} pour une explication. On rencontre parfois $\nu = 2$. Jamais plus. Le praticien choisit $\alpha \in \mathbb{R}$ tel que les trois termes de \eqref{ultralocal} aient le même ordre de grandeur. Il en découle qu'une identification précise du paramètre $\alpha$ est sans objet.

\subsubsection{PID intelligents}
Reprenons \eqref{ultralocal} avec $\nu = 2$. Un \og{PID intelligent}\fg, ou \og{iPID}\fg, est défini par
\begin{equation}\label{ipid}
\boxed{u = - \frac{F_{\text{est}} - \ddot{y}^\ast - K_P e - K_I \int e  - K_D \dot{e}}{\alpha}}
\end{equation}
où
\begin{itemize}
\item $y^\ast$ est la trajectoire de référence;
\item $e = y - y^\ast$ est l'erreur de poursuite;
\item $K_P, K_I, K_D \in \mathbb{R}$ sont les gains;
\item $F_{\text{est}}$ est une estimée de $F$.
\end{itemize}
Il vient
\begin{equation}\label{err2}
\ddot{e} = K_P e + K_I \int e + K_D \dot{e}  + F - F_{\text{est}} 
\end{equation}
Si l'estimée $F_{\text{est}}$ est \og{bonne}\fg, c'est-à-dire $F - F_{\text{est}} \backsimeq 0$, déterminer les gains est immédiat pour obtenir une \og{bonne}\fg \ poursuite. C'est un avantage important par rapport aux PID classiques.

Si $K_D = 0$, on a un \og{PI intelligent}\fg, ou \og{iPI}\fg,
\begin{equation}\label{ipi}
\boxed{u = - \frac{F_{\text{est}} - \ddot{y}^\ast - K_P e - K_I \int e}{\alpha}}
\end{equation}
Si $K_I = 0$, on a un \og{PD intelligent}\fg, ou \og{iPD}\fg,
\begin{equation}\label{ipd}
\boxed{u = - \frac{F_{\text{est}} - \ddot{y}^\ast - K_P e - K_D \dot{e}}{\alpha}}
\end{equation}
Avec $\nu = 1$, on aboutit à un \og{P intelligent}\fg, ou \og{iP}\fg, 
\begin{equation}\label{ip}
\boxed{u = - \frac{F_{\text{est}} - \ddot{y}^\ast - K_P e}{\alpha}}
\end{equation}
C'est le correcteur le plus courant dans la pratique du MFC\footnote{Voir, à ce sujet, \cite{ijc13} et \cite{iste}.}. La dynamique \eqref{err2} devient
$$
\dot{e} = K_P e  + F - F_{\text{est}} 
$$
\begin{remark}
On démontre \cite{andr,ijc13} qu'un iP (resp. iPD) est équivalent, en un certain sens, à un PI (resp. PID) classique. On comprend ainsi, sans doute pour la première fois, 
\begin{itemize}
\item l'universalité industrielle des PID,
\item la difficulté de leur réglage.
\end{itemize}
\end{remark}
\begin{remark}\label{mimo}
Dans tous les exemples \emph{MIMO}\footnote{Acronyme de \emph{Multi-Input Multi-Output}.} concrets rencontrés jusqu'à présent, on a toujours pu se ramener à des systèmes de type \eqref{ultralocal} en parallèle ou en cascade. La mise en place des correcteurs intelligents reste donc facile (voir \cite{quad,toulon,ieee17})\footnote{Aucun exemple concret n'a été découvert jusqu'à présent où cette approche est prise en défaut. Le \og{découplage}\fg, qui a suscité une recherche aussi abondante que prolixe en automatique théorique, linéaire ou non, est mis en question. Est-il besoin de rappeler une conclusion analogue \cite{ijc13}  à propos de la \og robustesse \fg?}.
\end{remark}

\subsubsection{Estimation de $F$}\label{2}
On peut approcher toute fonction $[a, b] \rightarrow  \mathbb{R}$, $a, b \in \mathbb{R}$, $a < b$, intégrable, c'est-à-dire très générale, par une fonction constante par morceaux (voir, par exemple, \cite{godement}, \cite{rudin1}). Par utilisation de fenêtres temporelles glissantes, l'estimation de $F$ se ramène ainsi à celle d'une constante $\Phi$. On présente, pour rester court, les calculs dans le cas $\nu = 1$. Les règles du calcul opérationnel (voir, par exemple, \cite{erde}) permettent de rééecrire
\eqref{ultralocal} ainsi
$$
sY = \frac{\Phi}{s}+\alpha U +y(0)
$$
On élimine la condition initiale $y(0)$ en dérivant les deux membres par $\frac{d}{ds}$:
$$
Y + s\frac{dY}{ds}=-\frac{\Phi}{s^2}+\alpha \frac{dU}{ds}
$$
On atténue le bruit en multipliant les deux membres par $s^{-2}$. La correspondance entre $\frac{d}{ds}$ et le produit par $-t$ permet de revenir au domaine temporel:
\begin{equation}\label{integral1}
{\small \boxed{F_{\text{est}}(t)  =-\frac{6}{\tau^3}\int_{t-\tau}^t \left\lbrack (\tau -2\sigma)y(\sigma)+\alpha\sigma(\tau -\sigma)u(\sigma) \right\rbrack d\sigma} }
\end{equation} 
où 
\begin{itemize}
\item $\tau > 0$ est \og{petit}\fg \ par rapport à la constante de temps la plus rapide;
\item $F_{\text{est}}(t)$ est la valeur numérique constante sur l'intervalle d'intégration attribuée à $F$.
\end{itemize}
On obtient \eqref{integral1} en temps réel.

On estime $F$, aussi en temps réel, par une formule différente grâce à \eqref{ip}:
\begin{equation}\label{integral2}
\boxed{F_{\text{est}}(t) = \frac{1}{\tau}\left[\int_{t - \tau}^{t}\left(\dot{y}^{\star}-\alpha u
- K_P e \right) d\sigma \right] }
\end{equation}
Après échantillonnage, \eqref{integral1} et \eqref{integral2} deviennent des filtres linéaires.

\begin{remark}
Au début, on estimait $F$ en utilisant un dérivateur \cite{mboup}  de $y$.
\end{remark}
\begin{remark}
Voir \cite{larminat} et \cite{carillo} pour d'autres approches de l'estimation.
\end{remark}

\subsection{ADRC}
On se conforme à l'usage, contraire à celui du MFC, en présentant l'ADRC par une équation différentielle:
\begin{equation}\label{adrc0}
y^{(n)}(t) = f\left(t, y(t), \dot{y}(t), \dots, y^{(n - 1)}(t)\right) + w(t) + a(t,y(t)) u(t)
\end{equation}
où
\begin{itemize}
\item la fonction $a \neq 0$ est connue;
\item l'ordre de dérivation $n \geqslant 1$ l'est aussi;
\item la fonction $f$, appelée, parfois, \og{perturbation interne}\fg, ou \og{endogène}\fg, l'est mal\footnote{Cet emploi du mot \og{perturbation}\fg \ pour désigner la structure mathématique du système nous semble malheureux.};
\item $w$ est une \og{perturbation externe}\fg, ou \og{exogène}\fg.
\end{itemize}
La some $f + w$ est la \og{perturbation totale}\fg. 

En ignorant $w$, \eqref{adrc0} devient:
\begin{equation}\label{adrc1}
y^{(n)}(t) = f\left(t, y(t), \dot{y}(t), \dots, y^{(n - 1)}(t)\right) + a(t,y(t)) u(t)
\end{equation}
\cite{ibero} a observé que \eqref{adrc1} vérifie la propriété de platitude et que la sortie $y$ est plate\footnote{C'est dire, rappelons-le, que $u$ s'exprime comme fonction de $y$ et de ses dérivées jusqu'à un ordre fini. On connaît la richesse de la littérature sur la platitude: voir, par exemple, \cite{murray}, \cite{levine}, \cite{riga}, \cite{tarbes}, \cite{rudolph}, \cite{sira1}, et leurs bibliographies.}. Si l'on emploie un \og{correcteur proportionnel intégral généralisé}\fg \ \cite{gpi}, ou \og{GPI}\fg, le \og{filtrage plat}\fg \  (\emph{flat filtering}),
\cite{sira4}, ou les techniques algébriques d'estimation (\cite{identif1,identif2,identif3},  et \\ \cite{sira2}),  l'approche de l'ADRC s'en trouve d'autant plus simplifiée \\ (\cite{cortes,sira3}). La distance considérable entre la voie prise par Sira-Ram\'{\i}rez, ses collaborateurs et les autres protagonistes de l'ADRC (voir, en particulier, le livre \cite{guo}) rend messéant, on en conviendra, un résumé plus complet.

\section{\'Eléments de comparaison}

\subsection{Prolégomènes}
Posons $\mathfrak{F} = f + w$. Avec cette notation de la perturbation totale, \eqref{adrc0} devient
\begin{equation}\label{adrc2}
y^{(n)} = \mathfrak{F}  + a u(t)
\end{equation}
à comparer à \eqref{ultralocal}, si l'on suppose $a$ constant, propriété usuelle, du reste, dans la présentation de l'ADRC:
\begin{enumerate}
\item la platitude ne joue aucun rôle en MFC, alors qu'elle est essentielle en ADRC avec, en plus, la nécessité pour $y$ en \eqref{adrc1} d'être une sortie plate\footnote{Ce fait est trop souvent ignoré par les tenants de l'ADRC.};
\item en cas de non-platitude, l'ADRC  doit, par, exemple se satisfaire du système linéaire variationnel, ou tangent, qui est plat s'il est commandable (voir \cite{furuta}, \\ \cite{sira3} pour le pendule dit, souvent, de Furuta);
\item $a$ en \eqref{adrc0}-\eqref{adrc1}-\eqref{adrc2} est connu, alors qu'$\alpha$ en \eqref{ultralocal} est ajusté approximativement par l'opérateur\footnote{De façon plus générale, la machinerie pour l'estimation et l'identification semblent moins simple avec l'ARDC.};
\item l'ordre $n$ de dérivation en \eqref{adrc0}-\eqref{adrc1}-\eqref{adrc2} est connu et, donc, imposé, alors que $\nu$ en \eqref{ultralocal} est choisi par l'opérateur avec, toujours jusqu'à présent, une valeur faible, $1$ ou $2$.
\end{enumerate}
D'où la sélection d'exemples ci-dessous: ils n'entrent pas dans le cadre \textit{a priori} de l'ADRC. 

\subsection{Un système linéaire}

\begin{figure}[htbp]
\centerline{\includegraphics[width=5.19in,height=2.83in]{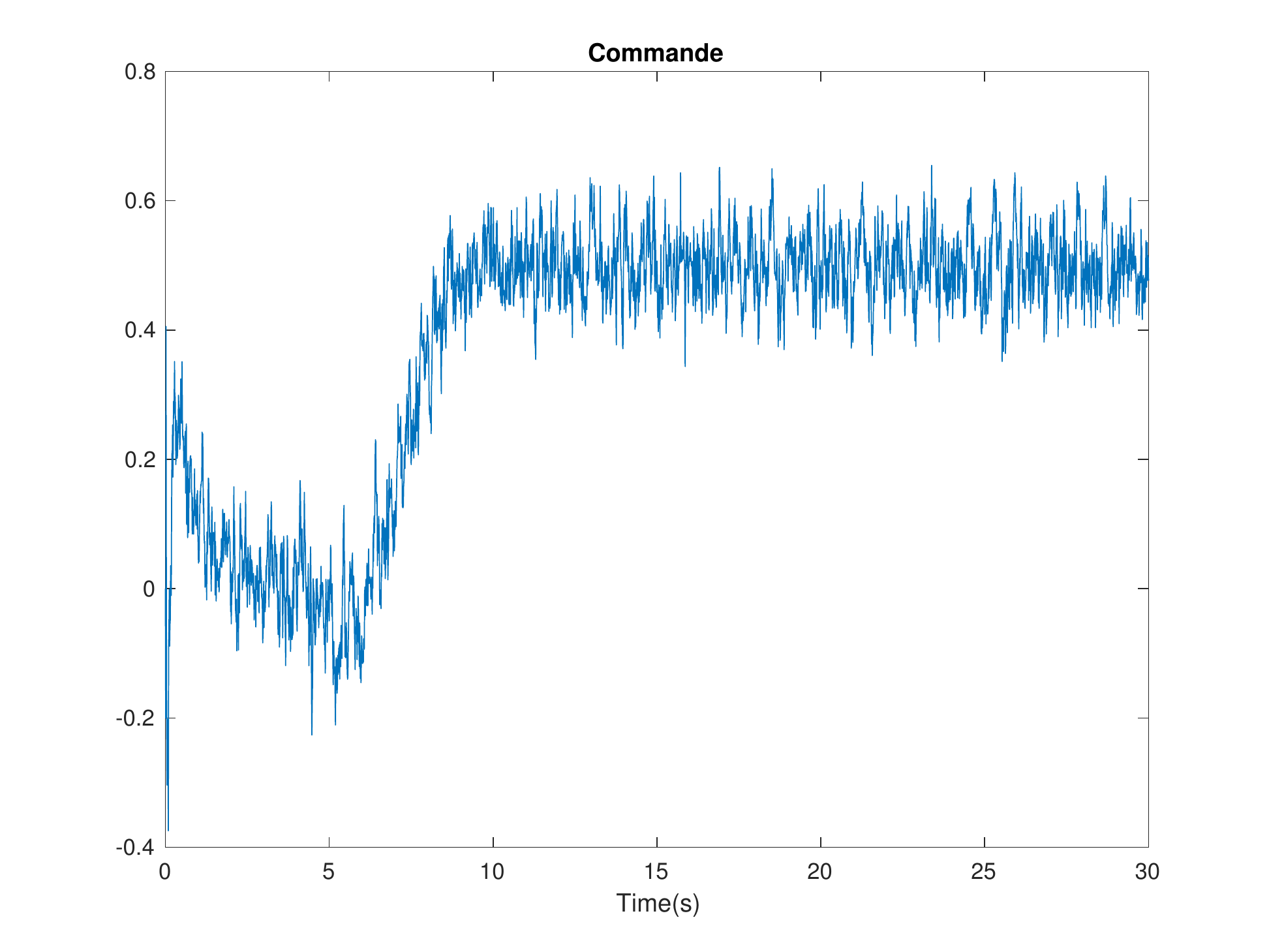}}
\caption{\textit{\sffamily Commande }}
\label{L1u}
\end{figure}

\begin{figure}[htbp]
\centerline{\includegraphics[width=5.19in,height=2.83in]{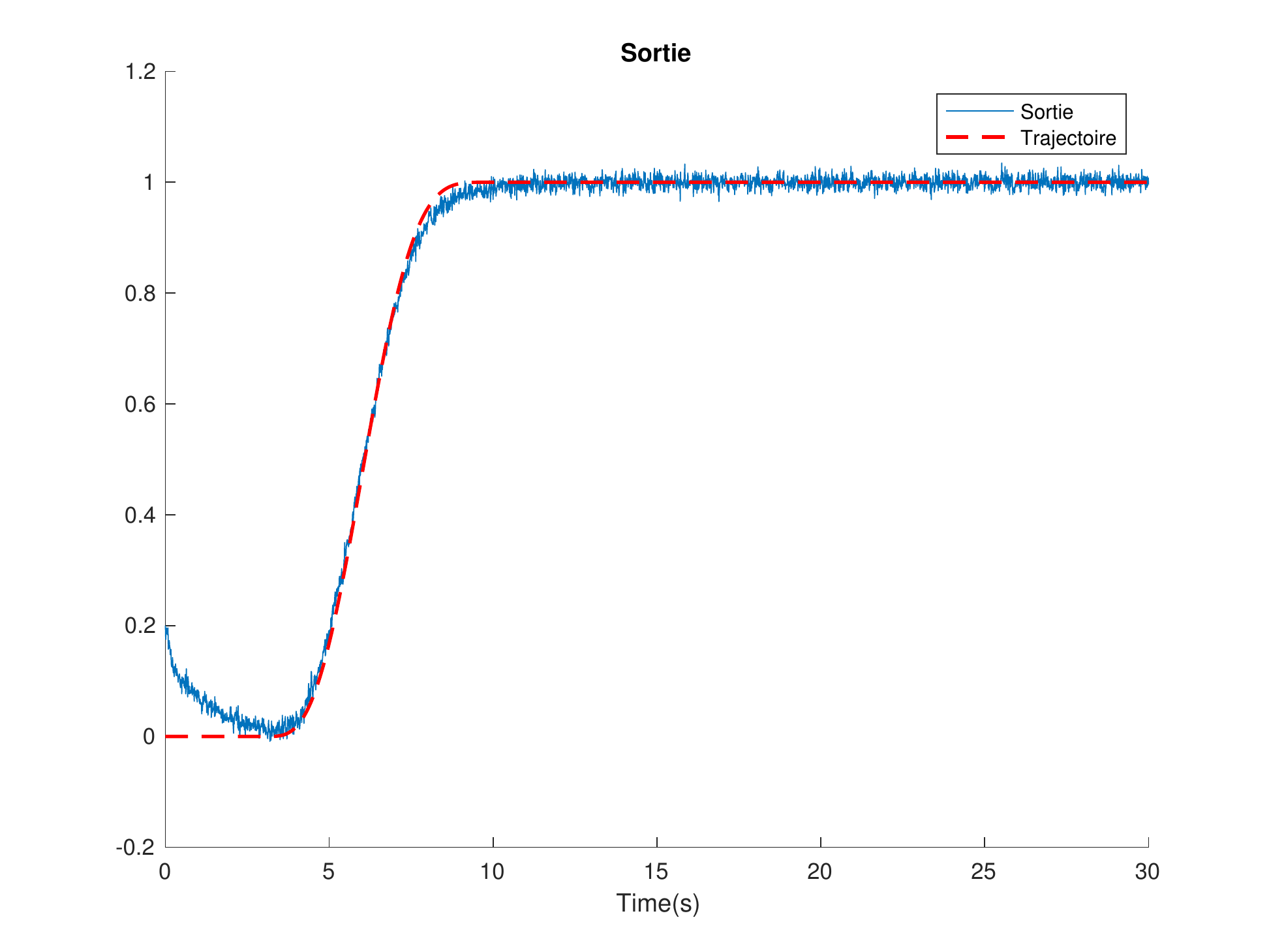}}
\caption{\textit{\sffamily Sortie }}
\label{L1y}
\end{figure}

\begin{figure}[htbp]
\centerline{\includegraphics[width=5.19in,height=2.83in]{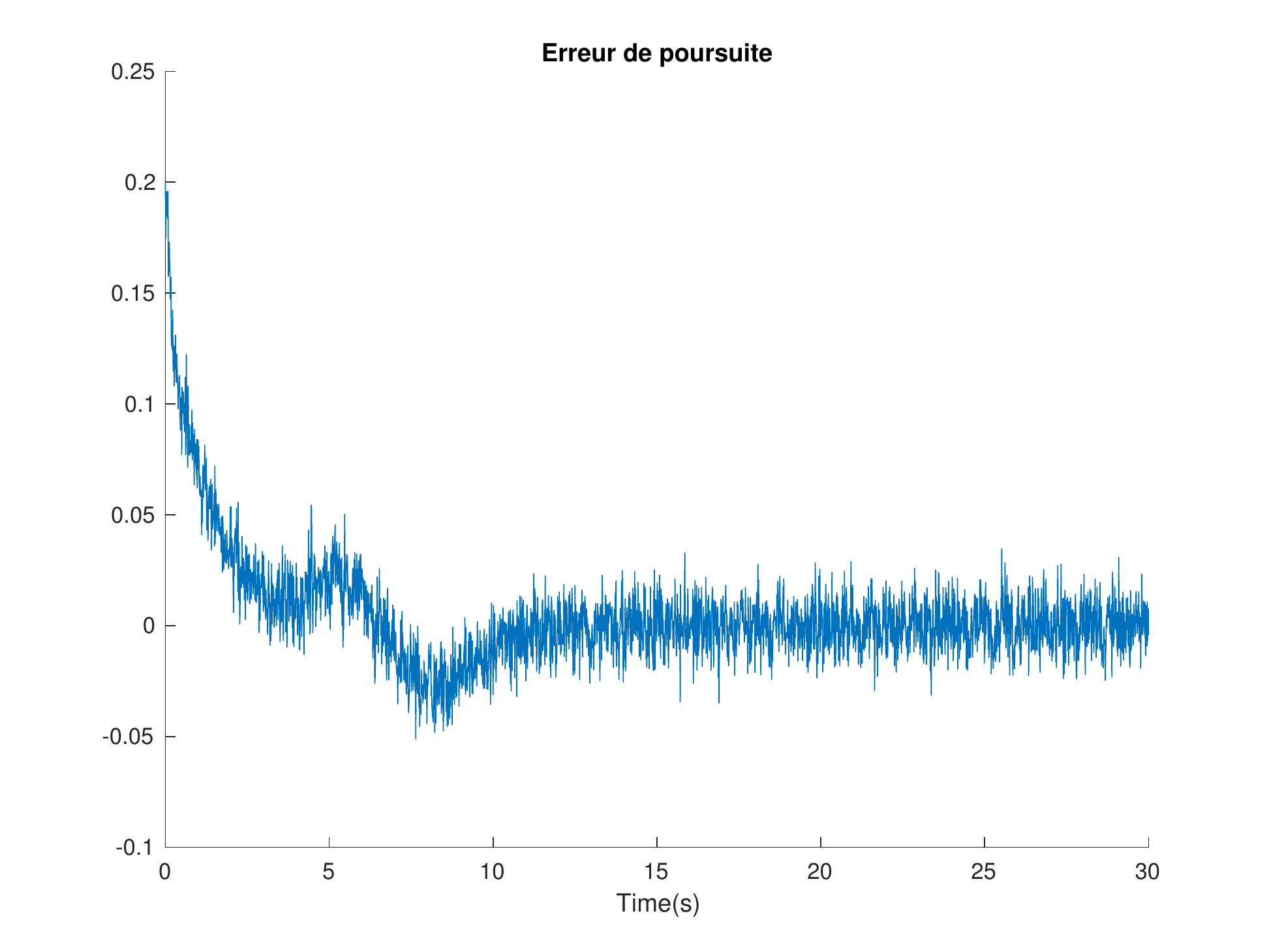}}
\caption{\textit{\sffamily Erreur de poursuite }}
\label{L1e}
\end{figure}

Pour un système linéaire stationnaire, de dimension finie, commandabilité et platitude sont équivalentes \cite{flmr1}. Pour un système SISO, la sortie $y$ est plate si, et seulement si, le numérateur de la fonction de transfert est constant \cite{lin}. Considérons donc le système défini par la fonction de transfert à numérateur non constant\footnote{Ce cas est d'autant plus éloquent qu'une présentation pour ingénieurs de l'ADRC \cite{herbst} ne considère que des fonctions de transfert à numérateurs constants.}:
$$
\frac{(s+2)^2}{(s-2)(s-1)(s+1)}
$$
Ici, un modèle ultralocal du premier ordre est utilisé avec $\alpha = K_p = 1$.
Un bruit additif, blanc, gaussien, centré, d'écart type 0.01, affecte la sortie. Les figures {\ref{L1u}} et {\ref{L1y}} présentent les résultats avec une période d'échantillonnage de $T_e=10$ms. La poursuite de la trajectoire de référence y est excellente comme le montre le tracé figure {\ref{L1e}}. Pour cet exemple, contrairement à ce qui suit, on représente la trajectoire par une courbe de Bézier \cite{Bez,abra} qui permet aisément des dérivées nulles en début et en fin.

\subsection{Cas non linéaires}

\subsubsection{Cas 1 \label{C1}}
\begin{figure}[htbp]
\centerline{\includegraphics[width=5.19in,height=2.83in]{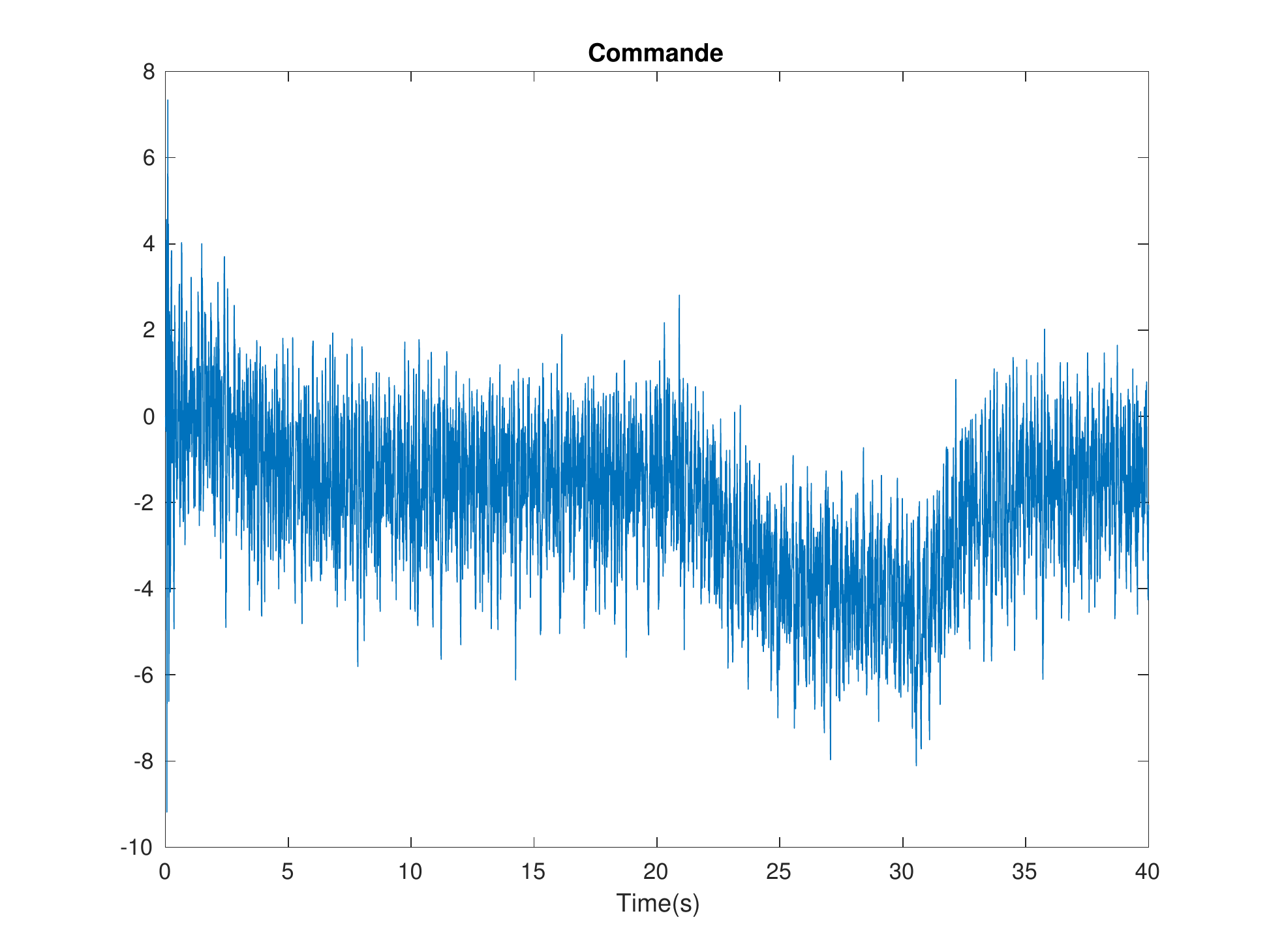}}
\caption{\textit{\sffamily Commande }}
\label{NL1u}
\end{figure}

\begin{figure}[htbp]
\centerline{\includegraphics[width=5.19in,height=2.83in]{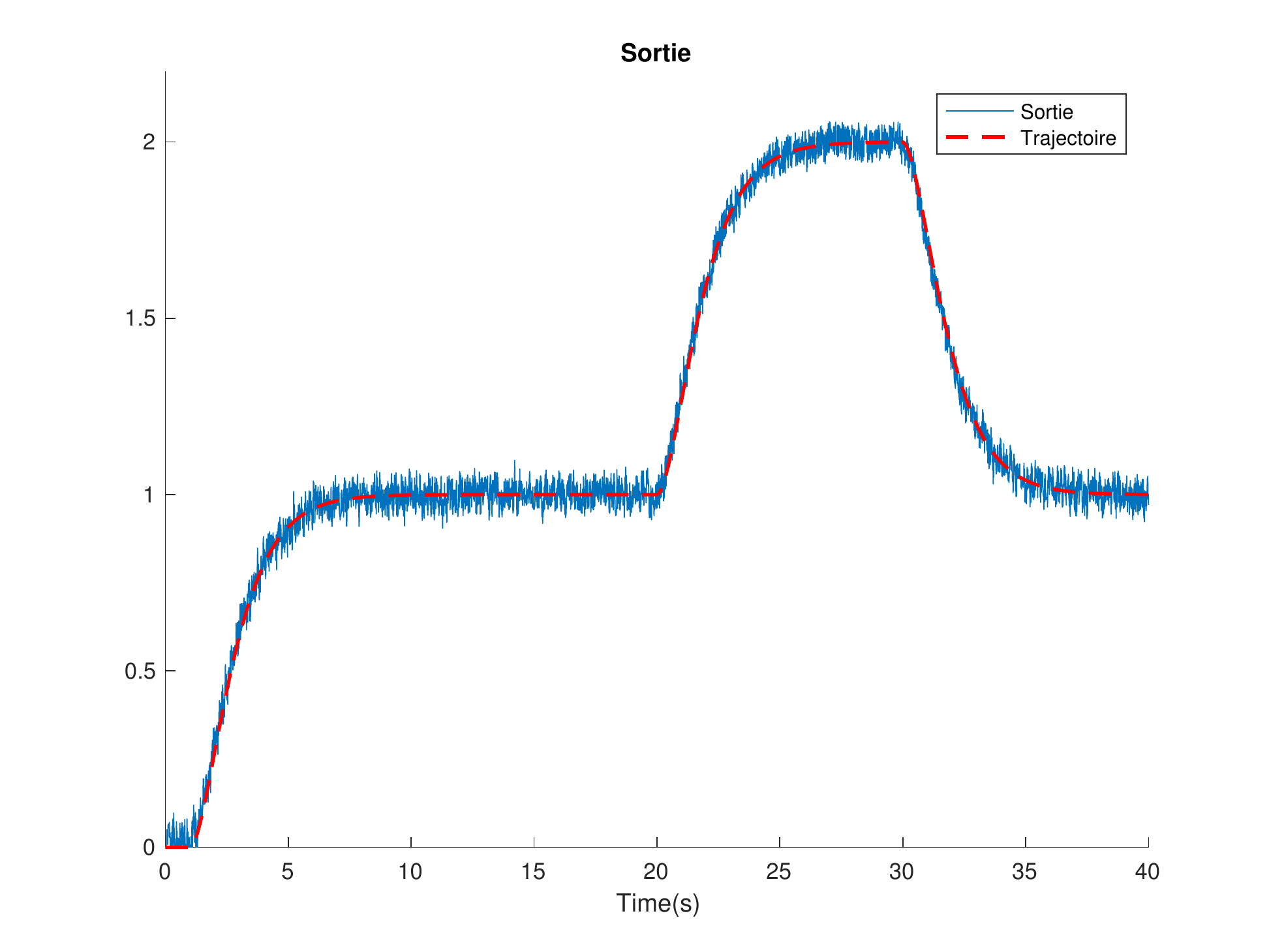}}
\caption{\textit{\sffamily Sortie }}
\label{NL1y}
\end{figure}

\begin{figure}[htbp]
\centerline{\includegraphics[width=5.19in,height=2.83in]{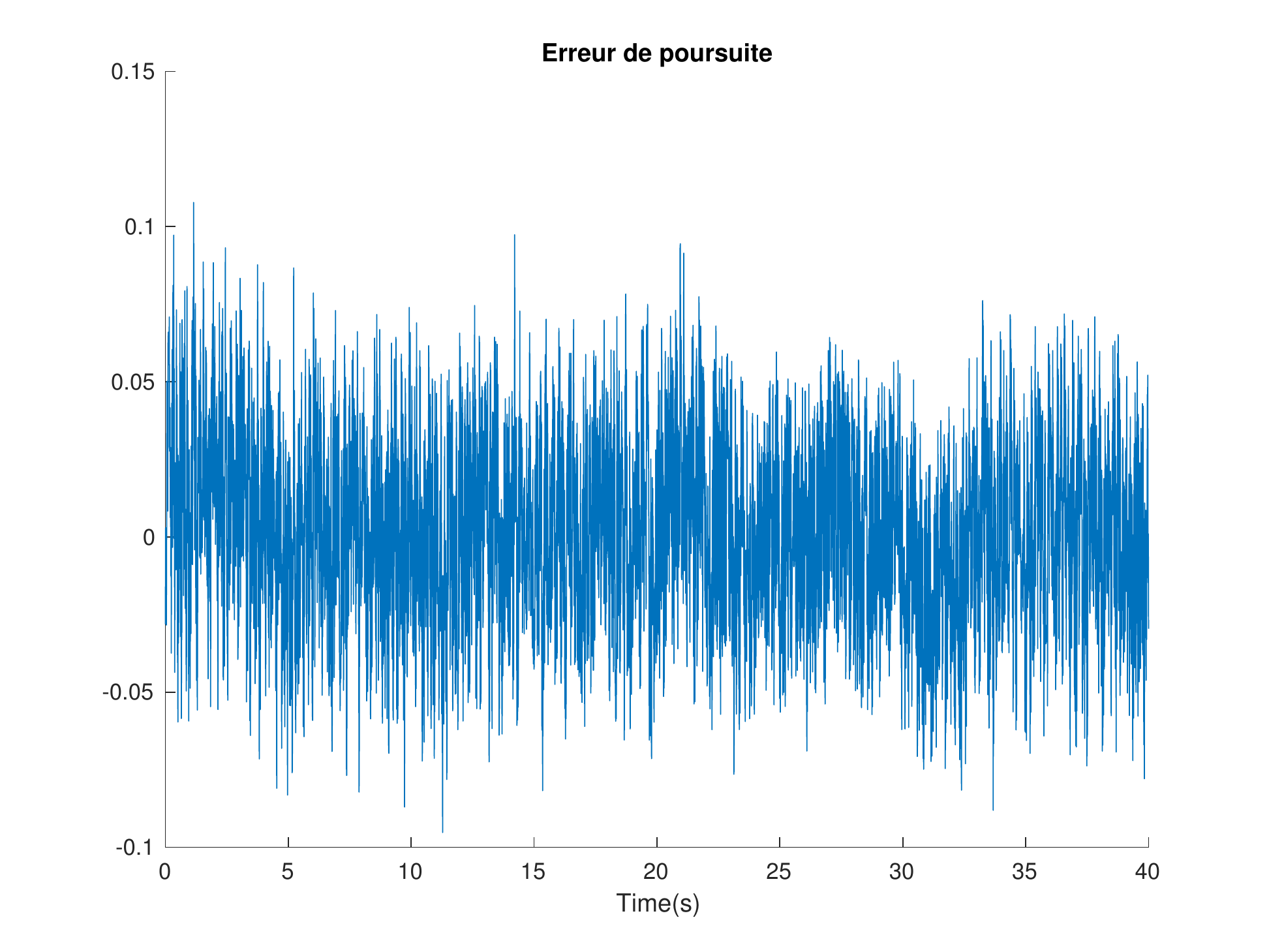}}
\caption{\textit{\sffamily Erreur de poursuite }}
\label{NL1e}
\end{figure}

Avec 
$$
\dot y-y= {\text{sign}} (u).\sqrt{|u|}
$$
on utilise \eqref{ultralocal} avec $\nu = 1$, $\alpha = 0.1$, $K_P = 1$. Un bruit additif, gaussien, centré, d'écart-type $0.05$, affecte la sortie. Avec une période d'échantillonnage $T_e = 10$ms, les simulations, reportées sur les figures 
{\ref{NL1u}}, {\ref{NL1y}} et {\ref{NL1e}}, attestent d'une poursuite de trajectoire irréprochable.


\subsubsection{Cas 2}
\begin{figure}[htbp]
\centerline{\includegraphics[width=5.19in,height=2.83in]{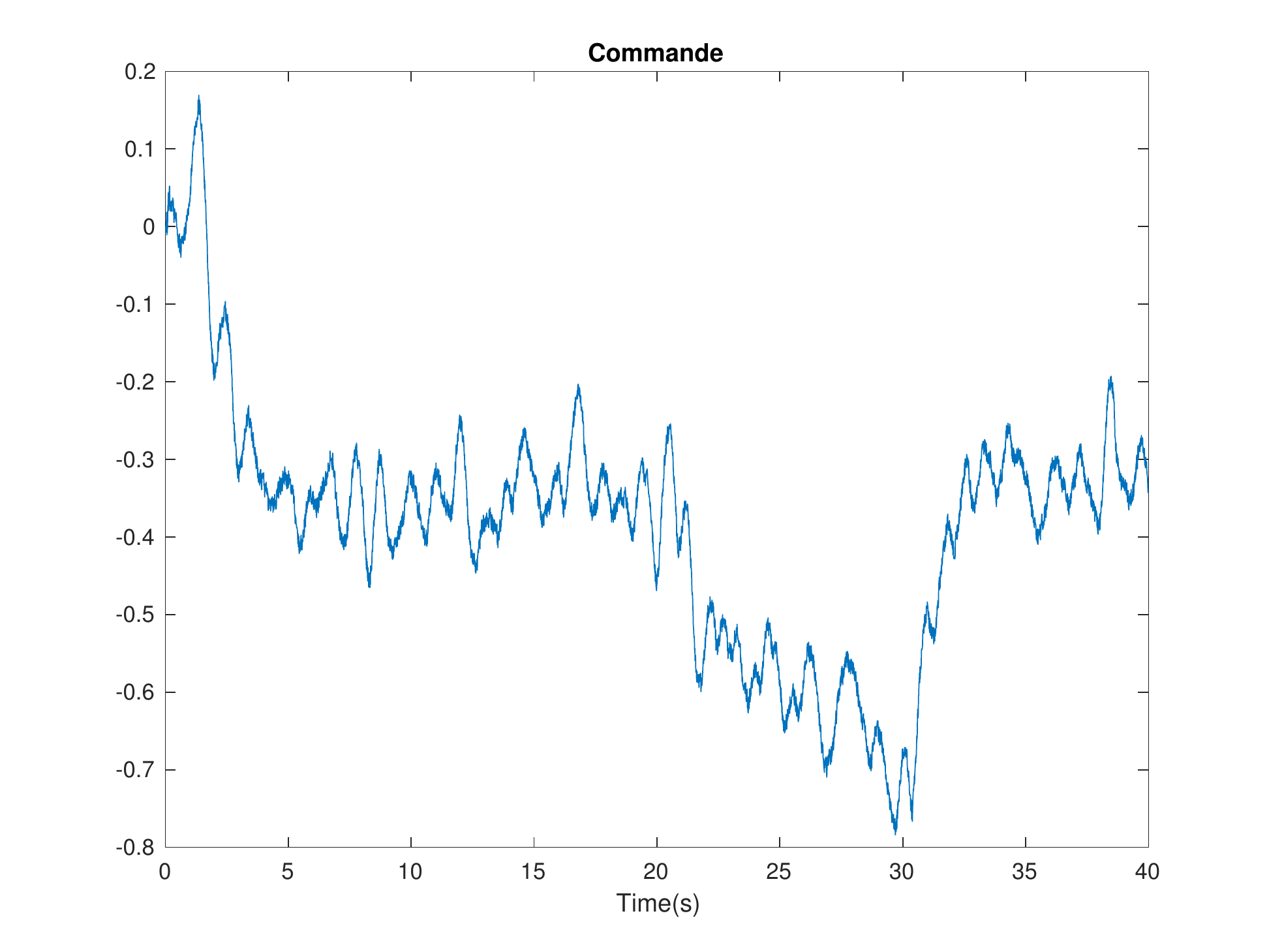}}
\caption{\textit{\sffamily Commande }}
\label{NL2u}
\end{figure}

\begin{figure}[htbp]
\centerline{\includegraphics[width=5.19in,height=2.83in]{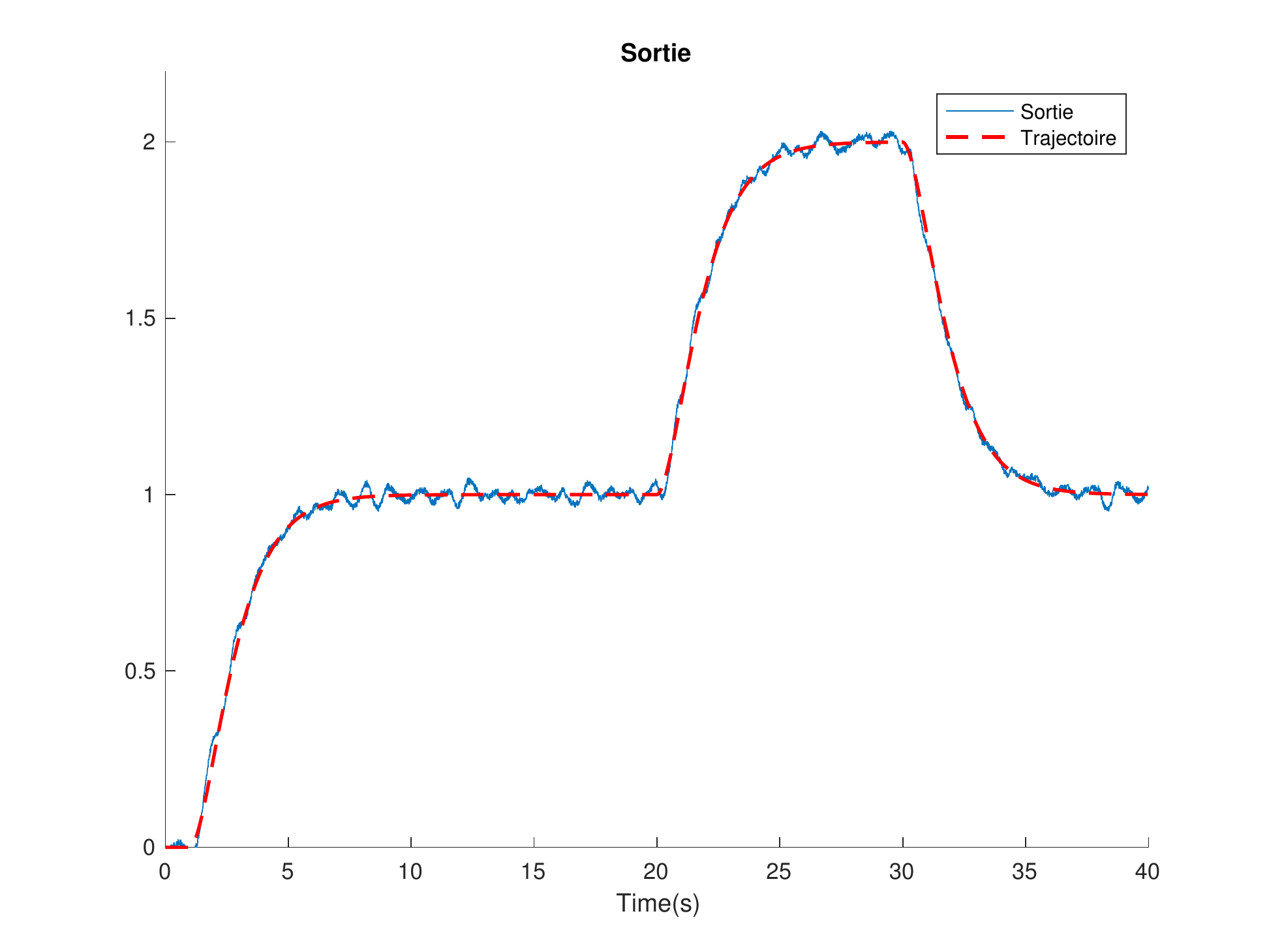}}
\caption{\textit{\sffamily Sortie }}
\label{NL2y}
\end{figure}

\begin{figure}[htbp]
\centerline{\includegraphics[width=5.19in,height=2.83in]{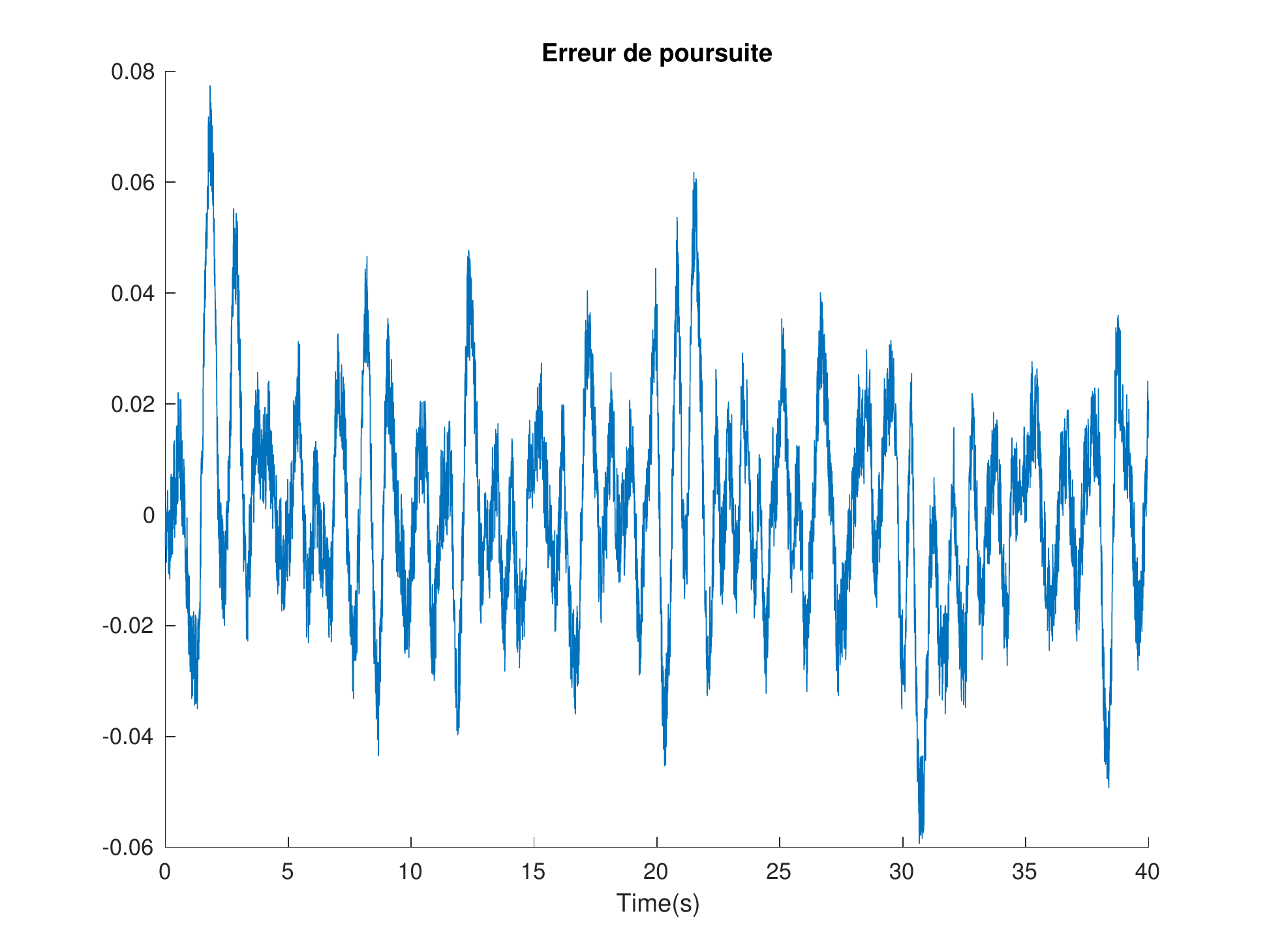}}
\caption{\textit{\sffamily Erreur de poursuite }}
\label{NL2e}
\end{figure}

Avec
$$\ddot y-1.5\dot y-y=(u+\dot u)^3$$
on choisit, pour \eqref{ultralocal}, $\nu = 1$, $\alpha = K_P = 10$. Avec un bruit sur $y$, additif, blanc, gaussien, centré, d'écart-type $0.01$, et un échantillonnage de période $T_e = 10$ms, la poursuite de trajectoire, exhibée par les figures 
{\ref{NL2u}}, {\ref{NL2y}} et {\ref{NL2e}} est bonne.

\subsubsection{Cas 3}
\begin{figure}[htbp]
\centerline{\includegraphics[width=5.19in,height=2.83in]{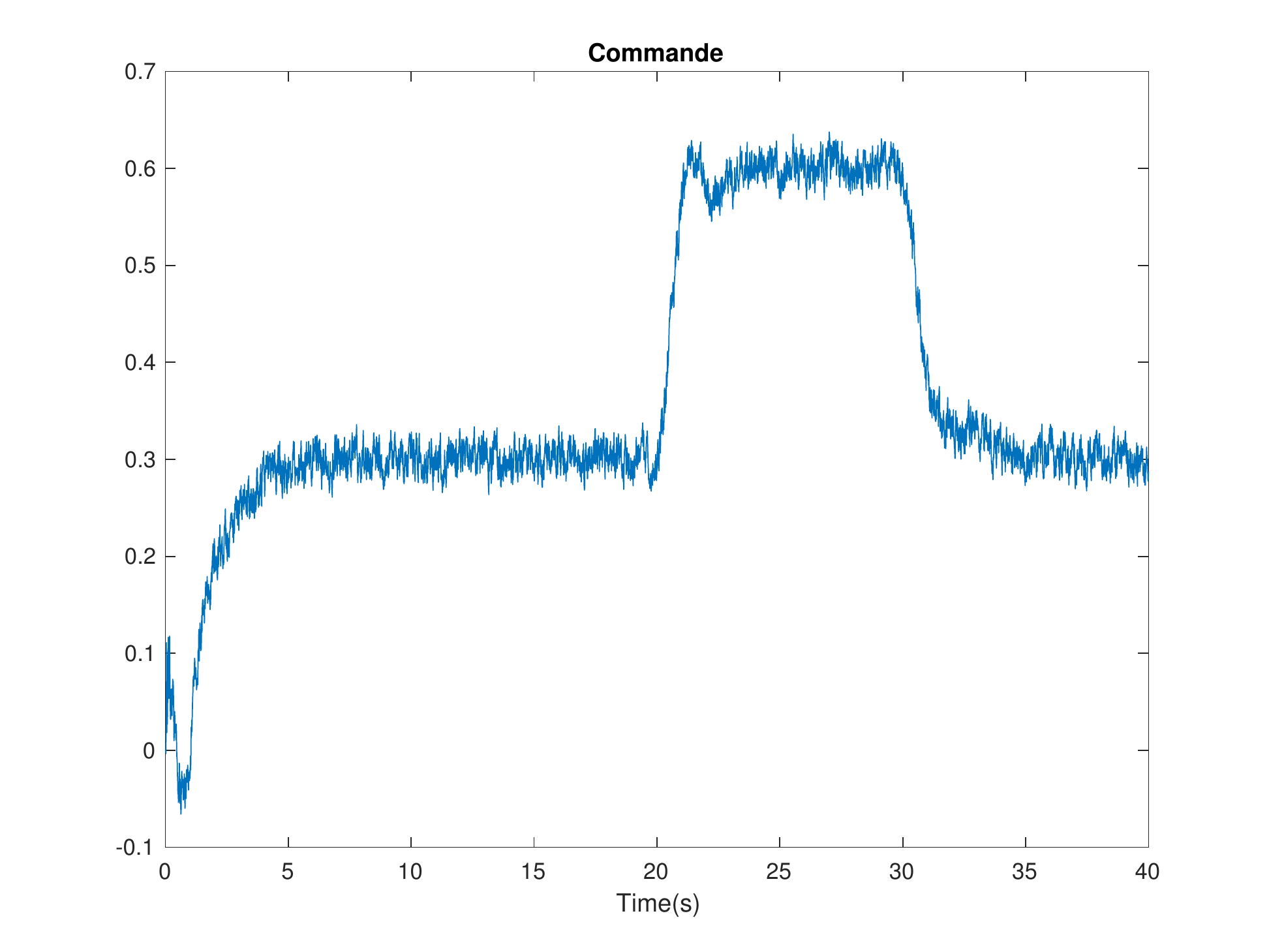}}
\caption{\textit{\sffamily Commande }}
\label{NL3u}
\end{figure}

\begin{figure}[htbp]
\centerline{\includegraphics[width=5.19in,height=2.83in]{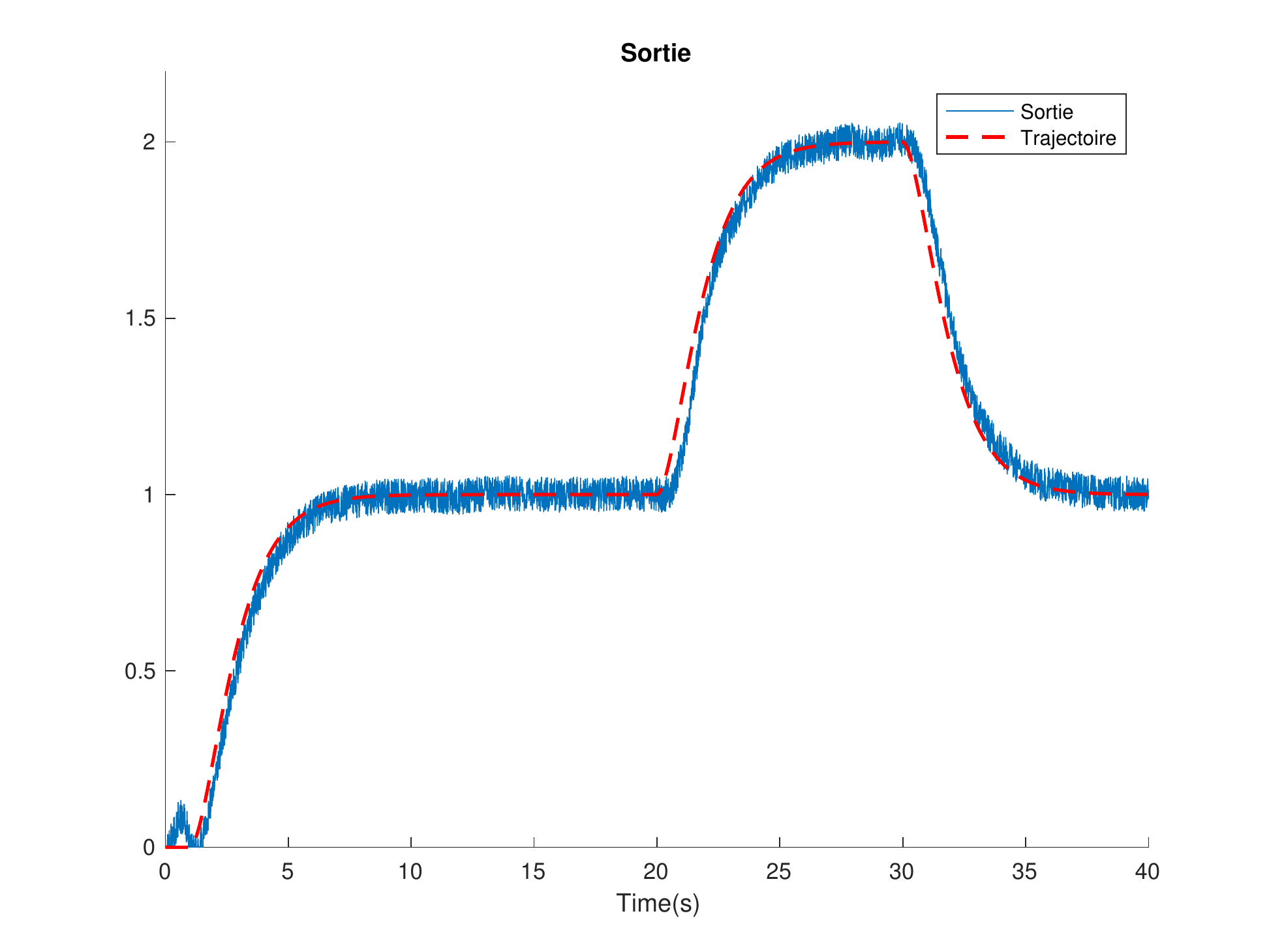}}
\caption{\textit{\sffamily Sortie }}
\label{NL3y}
\end{figure}

\begin{figure}[htbp]
\centerline{\includegraphics[width=5.19in,height=2.83in]{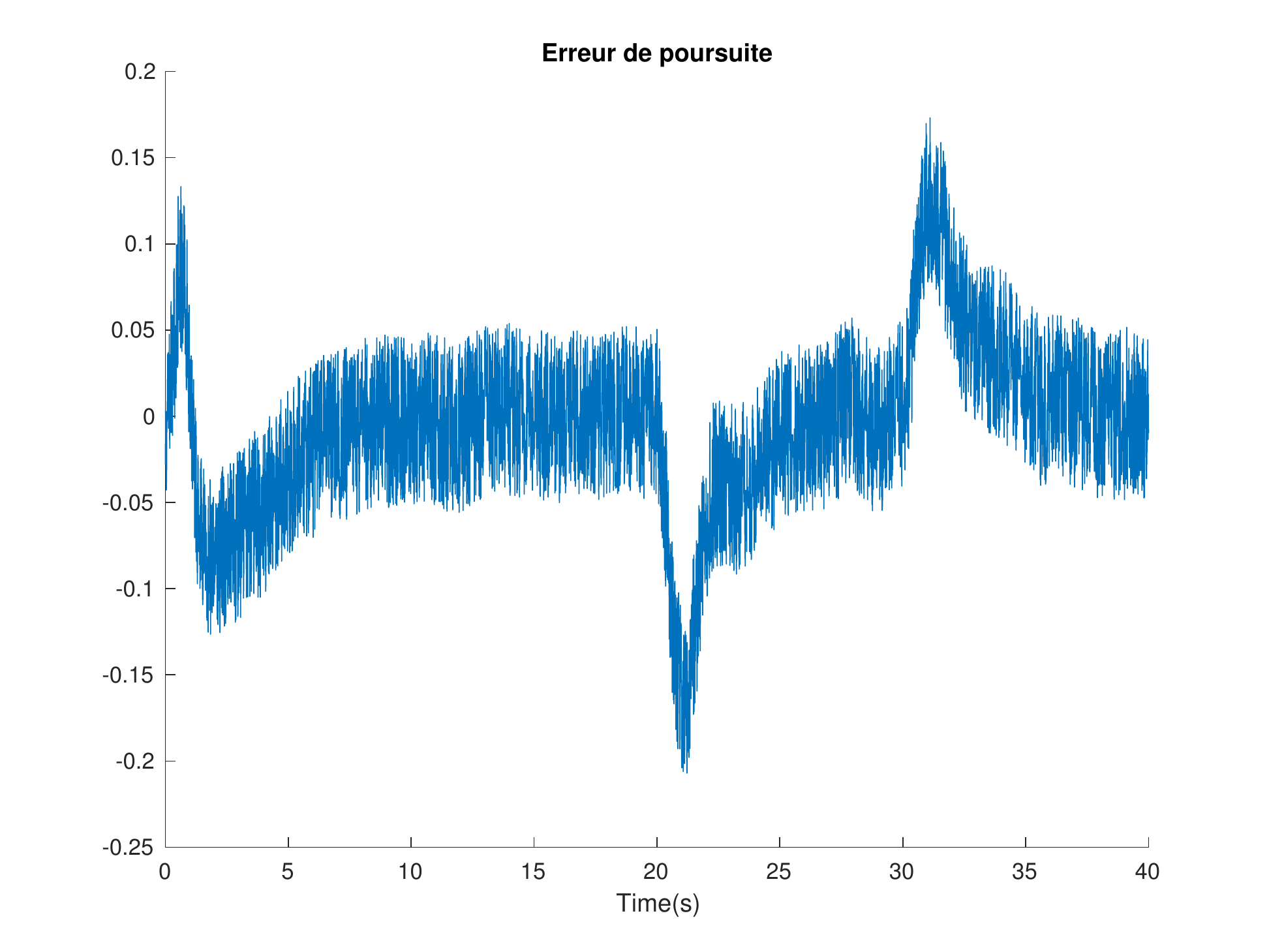}}
\caption{\textit{\sffamily Erreur de poursuite }}
\label{NL3e}
\end{figure}

Avec
$$
\ddot y+3\dot y+2y = \text{sign} (u).10^{|u|}
$$
$\nu = 1$, $\alpha = 10$, $K_P = 1$ en \eqref{ultralocal}. Les résultats des figures {\ref{NL3u}}, {\ref{NL3y}} et {\ref{NL3e}} sont excellents.



\subsubsection{Cas 4}
\begin{figure}[htbp]
\centerline{\includegraphics[width=5.19in,height=2.83in]{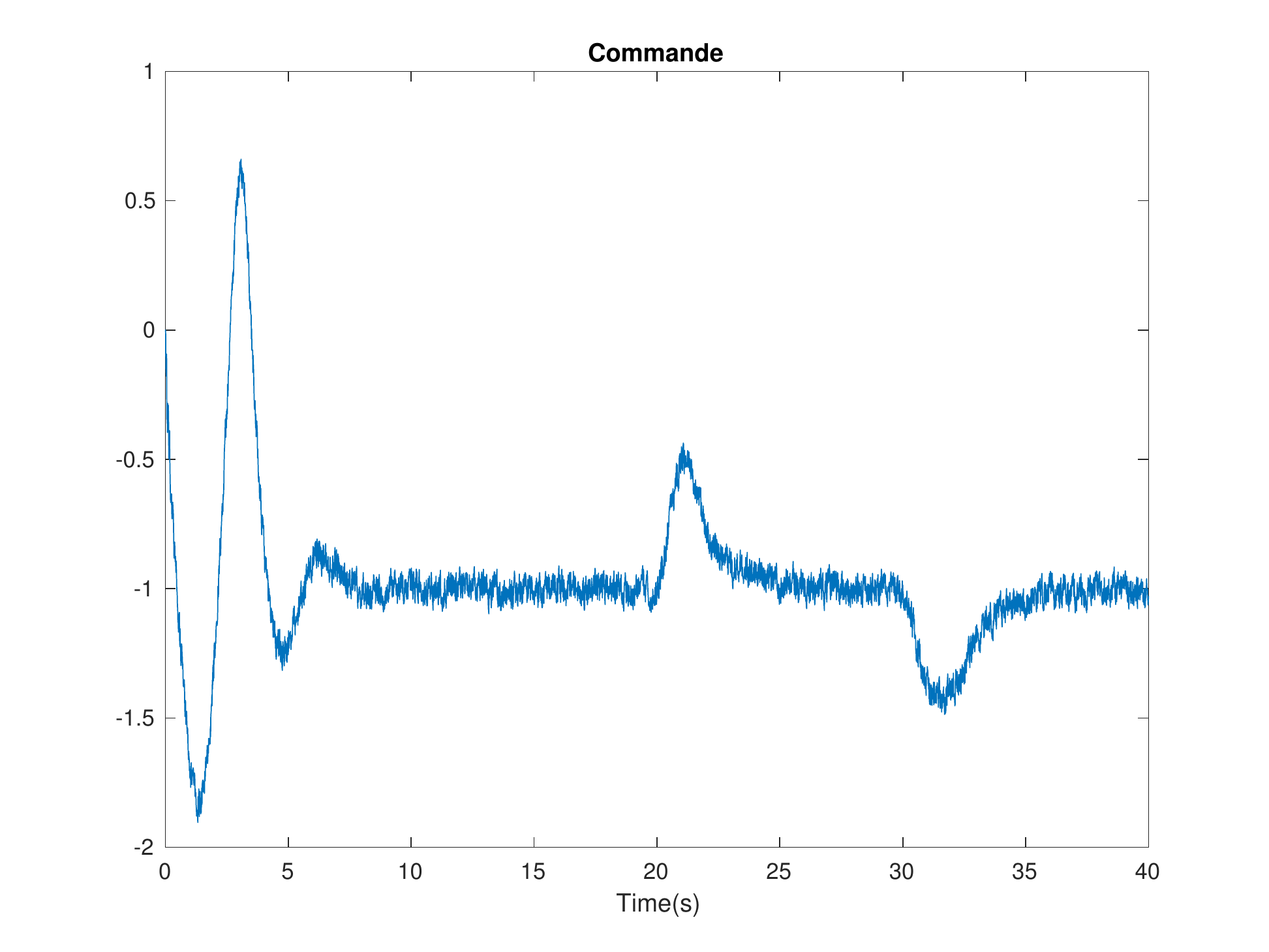}}
\caption{\textit{\sffamily Commande }}
\label{NL4u}
\end{figure}

\begin{figure}[htbp]
\centerline{\includegraphics[width=5.19in,height=2.83in]{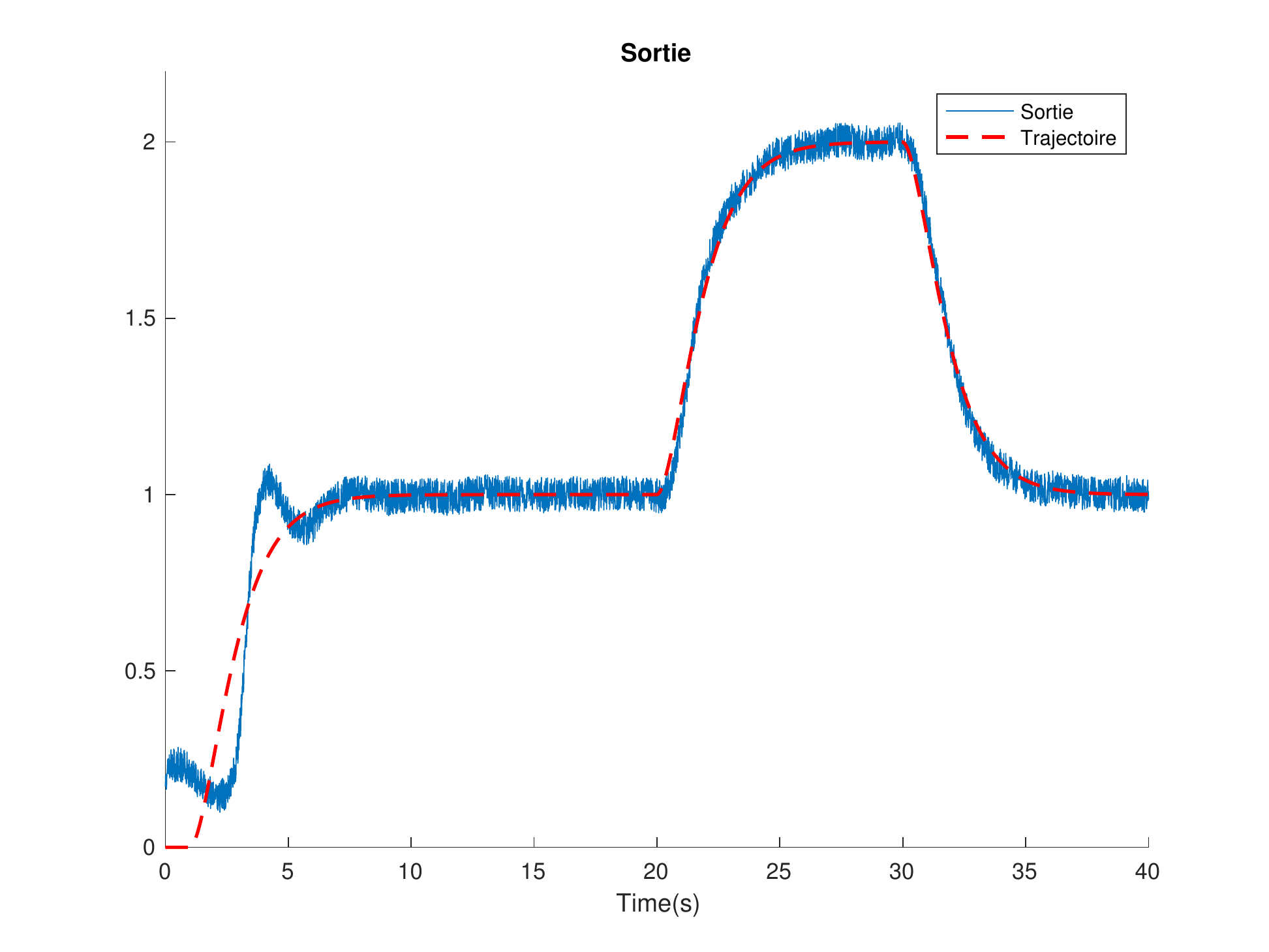}}
\caption{\textit{\sffamily Sortie }}
\label{NL4y}
\end{figure}

\begin{figure}[htbp]
\centerline{\includegraphics[width=5.19in,height=2.83in]{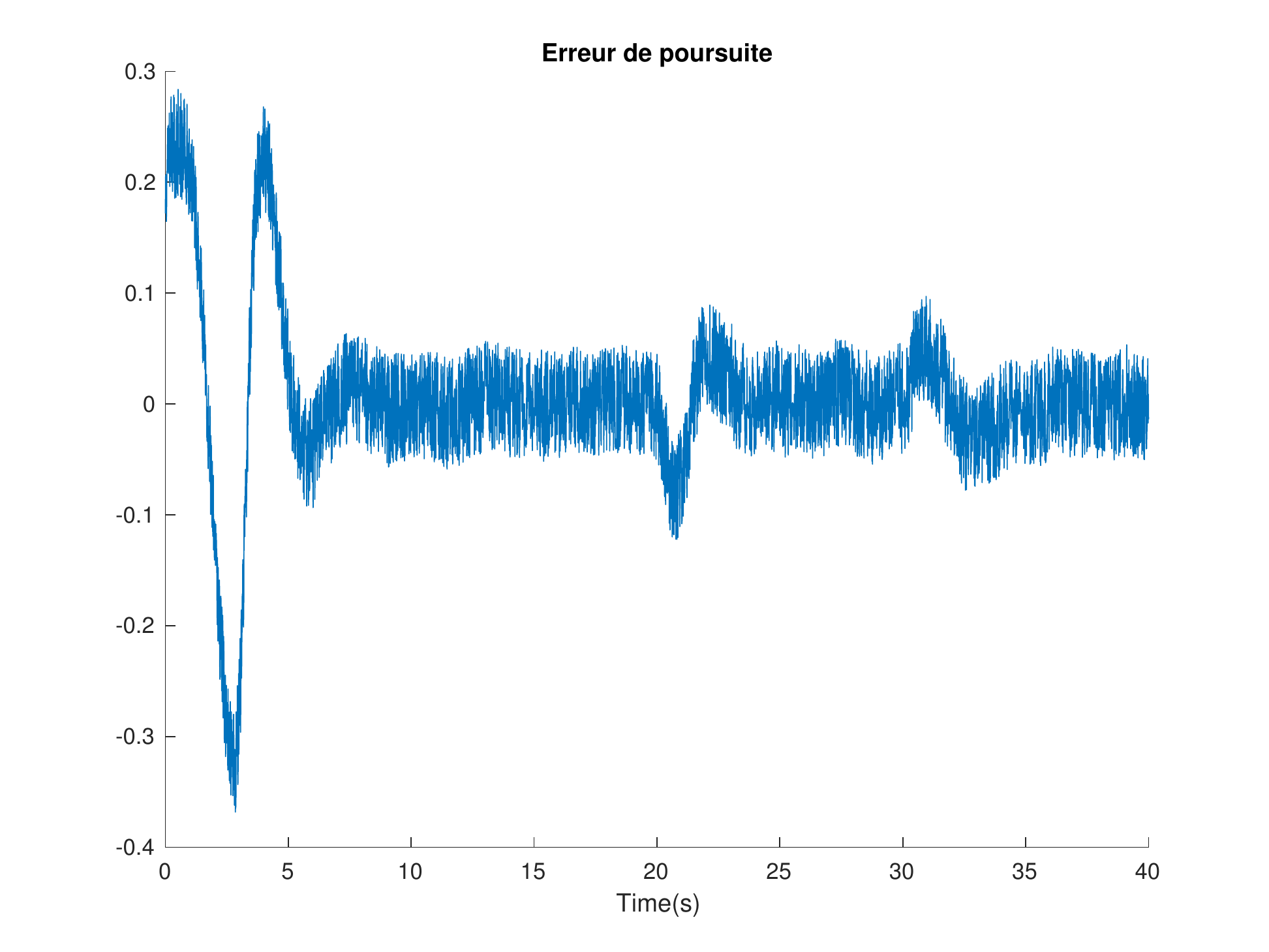}}
\caption{\textit{\sffamily Erreur de poursuite }}
\label{NL4e}
\end{figure}

Quoique le système
$$\dot y-y=(0.5\dot y+y)u$$
soit plat, de sortie plate $y$, il est, en raison du produit $\dot{y} u$ et comme on le voit en \eqref{adrc1}, plus général que ceux considérés en ADRC. De plus, son instabilité le rend délicat à commander. En \eqref{ultralocal}, on choisit $\nu = 1$, $\alpha = 5$, $K_P = 3$. On ajoute un bruit additif sur la sortie, blanc, gaussien, centré, d'écart-type $0.05$. La condition initiale est $y(0)=0.2$. La période d'échantillonnage est $T_e=10$ms. Les résultats de poursuite de trajectoire reportés sur les figures {\ref{NL4u}}, {\ref{NL4y}} et {\ref{NL4e}} sont tout à fait corrects.


\subsubsection{Cas 5}
\begin{figure}[!ht]
\centering

        \psfrag{p1}[cc][t]{\footnotesize pompe $1$}
        \psfrag{p2}[cc][t]{\footnotesize pompe $2$}
        \psfrag{q1}[l][cc]{\footnotesize$u_1$}
        \psfrag{q2}[r][cc]{\footnotesize$u_2$}
        \psfrag{c1}[r][cc]{\footnotesize cuve $1$}
        \psfrag{c2}[r][cc]{\footnotesize cuve $2$}
        \psfrag{c3}[r][cc]{\footnotesize cuve $3$}
        \psfrag{s}[l][cc]{$\tiny S$}
%
        \psfrag{q13}[cc][cc]{\tiny}
        \psfrag{q32}[cc][cc]{\tiny}
        \psfrag{q20}[cc][cc]{\tiny}
        \psfrag{m1}[l][cc]{\footnotesize$S_p,\mu_1$}
        \psfrag{m2}[l][cc]{\footnotesize$S_p,\mu_2$}
        \psfrag{m3}[l][cc]{\footnotesize$S_p,\mu_3$}
%
%
        \psfrag{l1}[r][cc]{\footnotesize$x_1$}
        \psfrag{l2}[r][cc]{\footnotesize$x_2$}
        \psfrag{l3}[r][cc]{\footnotesize$x_3$}
%
{\includegraphics[scale=0.6]{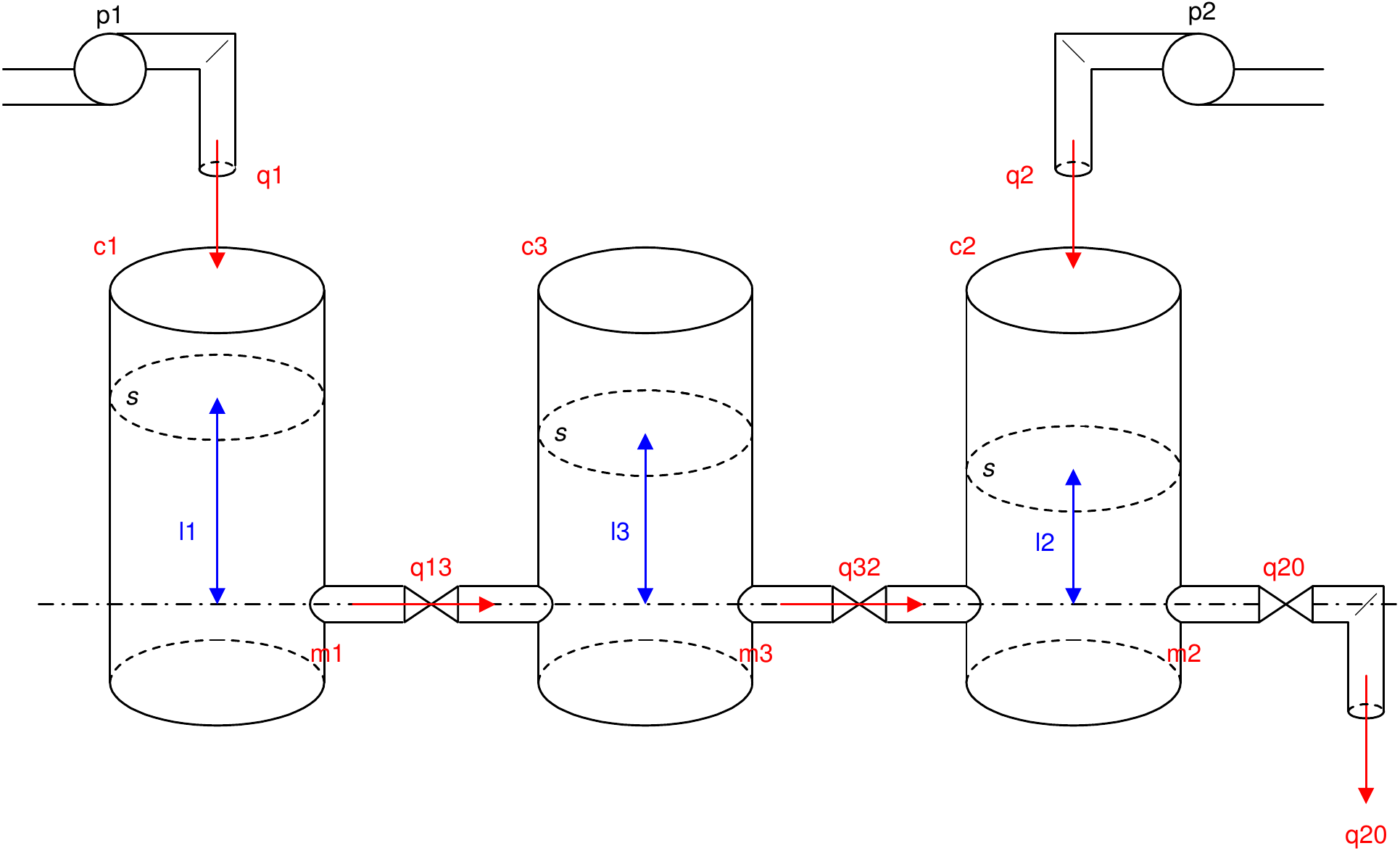}} \caption{Système des trois
cuves \label{schem}}
\end{figure}

\begin{figure}[htbp]
\centerline{\includegraphics[width=5.19in,height=2.83in]{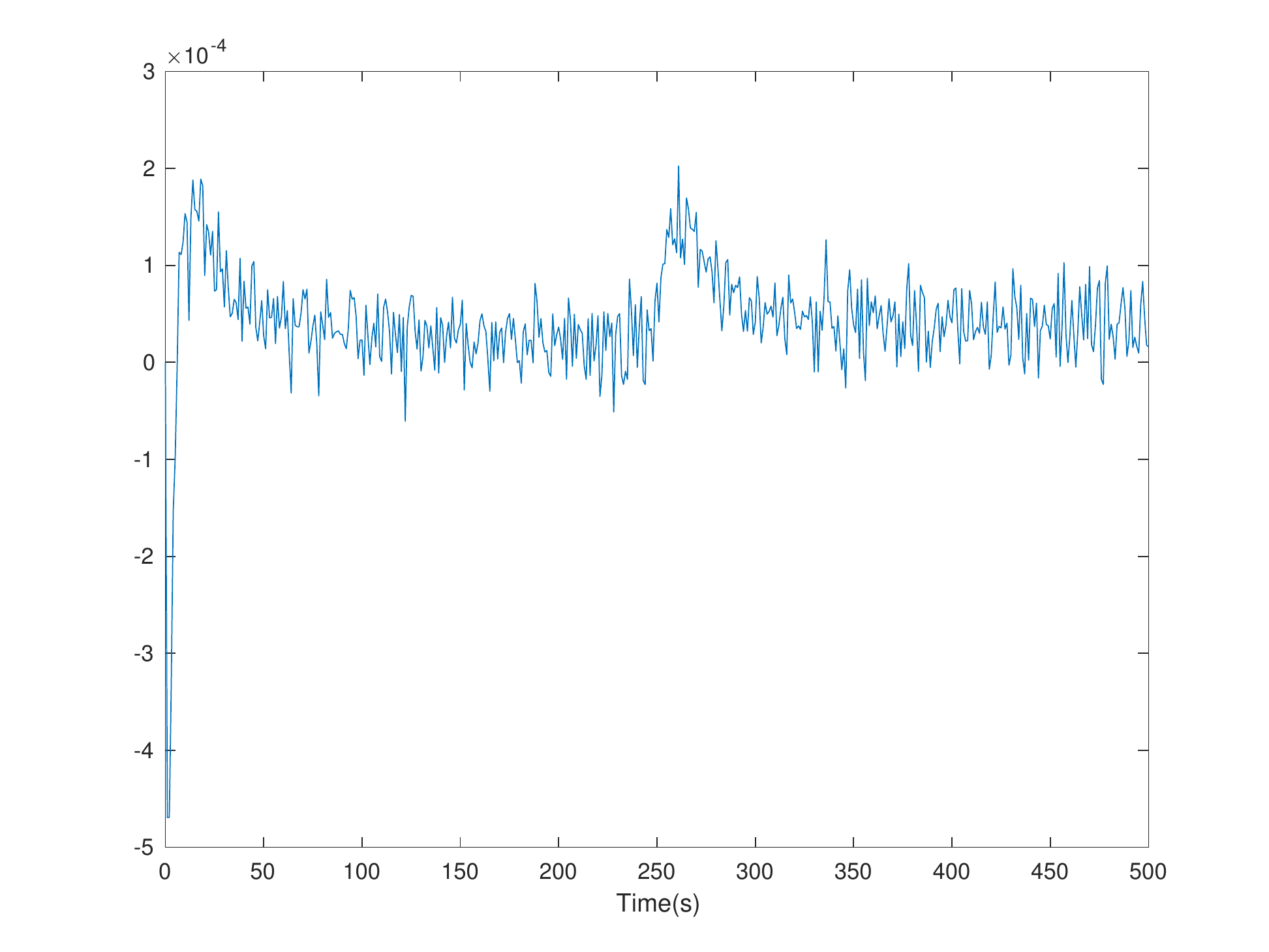}}
\caption{\textit{\sffamily Commande de la pompe 1}}
\label{C_u1}
\end{figure}

\begin{figure}[htbp]
\centerline{\includegraphics[width=5.19in,height=2.83in]{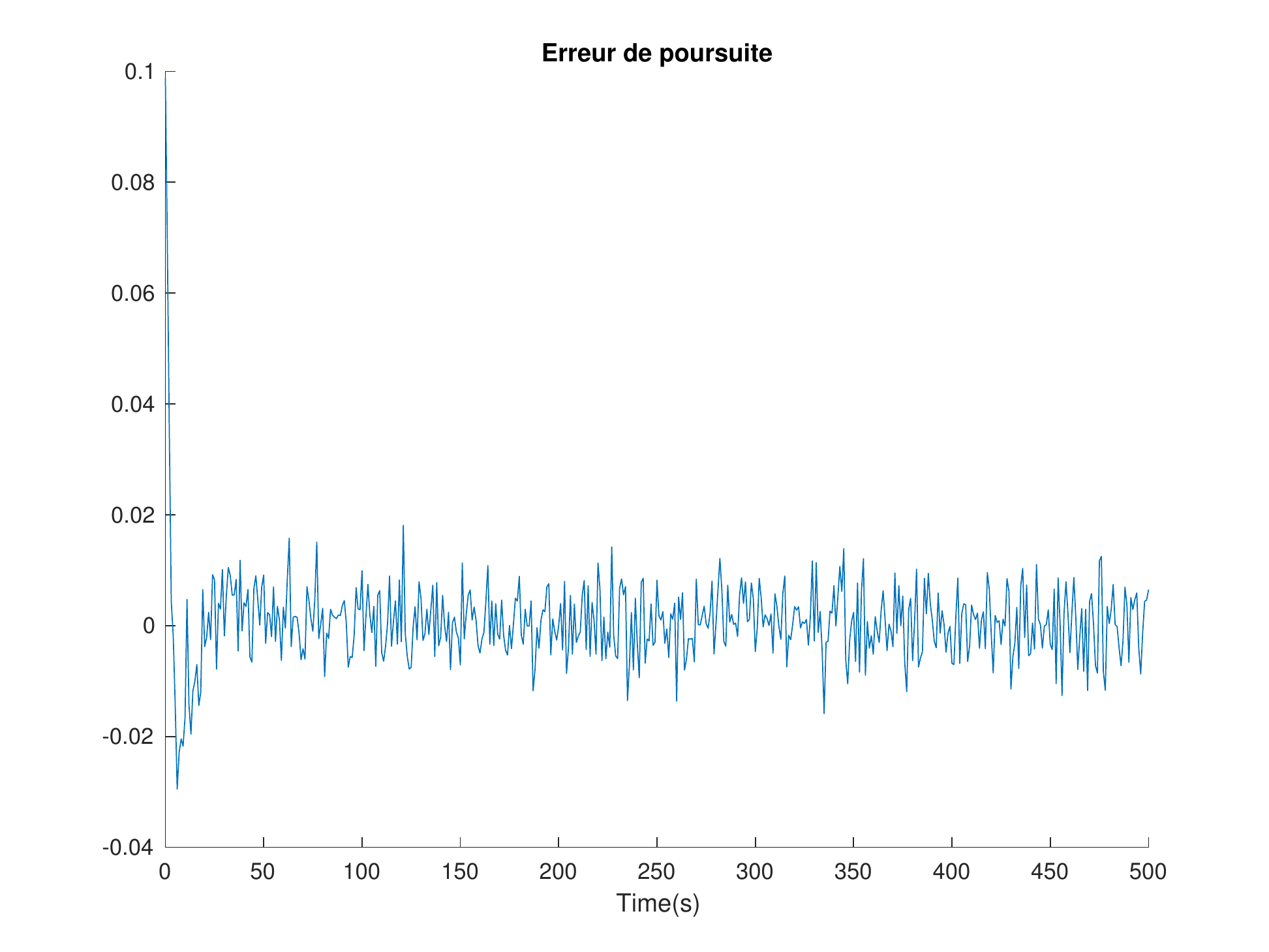}}
\caption{\textit{\sffamily Erreur de poursuite pour la cuve 1 }}
\label{C_e1}
\end{figure}

\begin{figure}[htbp]
\centerline{\includegraphics[width=5.19in,height=2.83in]{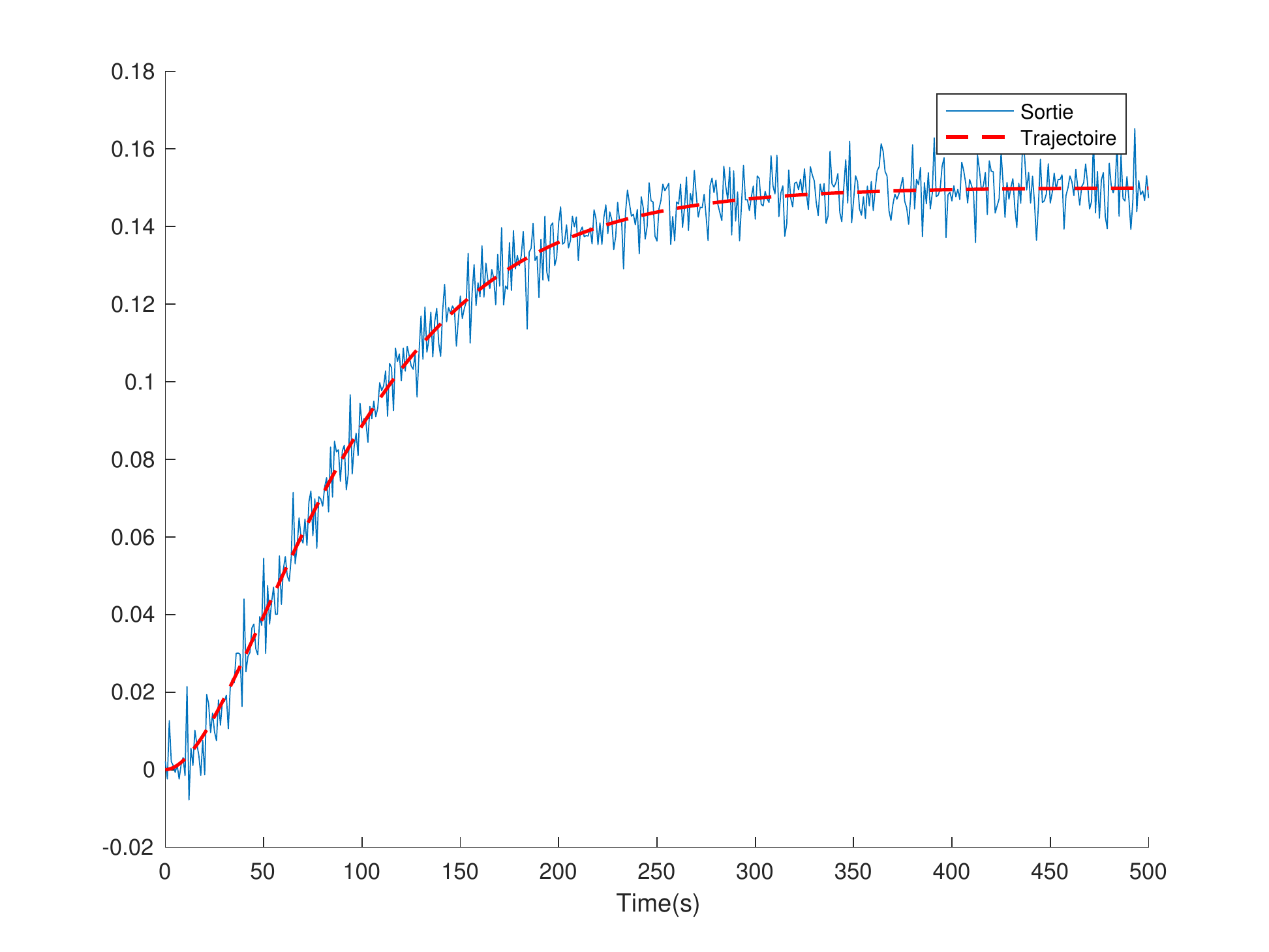}}
\caption{\textit{\sffamily Evolution du niveau dans la cuve 2}}
\label{C_l2}
\end{figure}

\begin{figure}[htbp]
\centerline{\includegraphics[width=5.19in,height=2.83in]{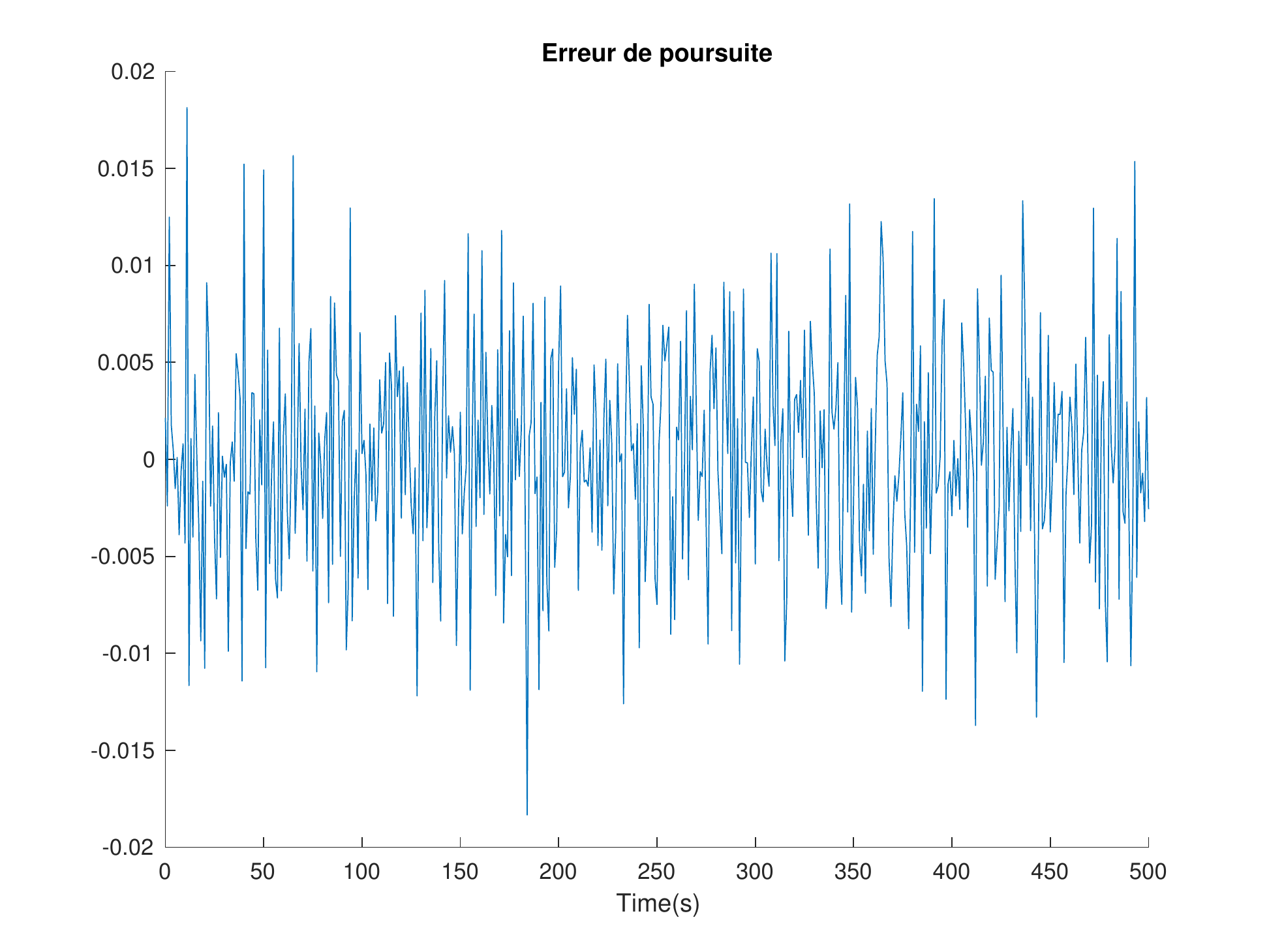}}
\caption{\textit{\sffamily Erreur de poursuite pour la cuve 2 }}
\label{C_e2}
\end{figure}

Une littérature considérable a été consacrée au système, plutôt académique, des trois cuves, notamment en diagnostic, représenté par la figure {\ref{schem}} (voir, par exemple, \cite{cran} et ses références).
Confirmant la remarque {\ref{mimo}}, il témoigne de la simplicité du traitement des systèmes multivariables par rapport à l'ADRC (voir, par exemple, \cite{guo}, \cite{inoue}, \cite{sira3}, et \cite{sun}). En voici la description par équations différentielles
\begin{equation}
    \begin{cases}\begin{array}{ll}
    \dot{x}_1=& - D \mu_1 \text{sign}(x_1 -x_3)\sqrt{|x_1-x_3|}+u_1/S\\
    \dot{x}_2=& D \mu_3 \text{sign}(x_3 - x_2)\sqrt{|x_3 - x_2|}-D \mu_2 \text{sign}(x_2)\sqrt{|x_2|}+u_2(t)/S\\
    \dot{x}_3 =& - D \mu_1 \text{sign}(x_1 - x_3)\sqrt{|x_1 - x_3|} D \mu_3 \text{sign}(x_3 - x_2)\sqrt{|x_3 - x_2|}\\
    y_{1}=&x_1  \\
    y_{2}=&x_2 \\
    y_{3}=&x_3 \\
    \end{array}\end{cases}\label{eq_3cuves}
\end{equation}
où 
\begin{itemize}
\item $x_i$, $i = 1, 2, 3$, est le niveau de liquide dans la cuve $i$, 
\item les commandes $u_1$, $u_2$ sont les arrivées de liquide, 
\item $D=\left(S_p\sqrt{2 g}\right)/S$, 
\item $g = 9.81 \text{m.s}^{-2}$ est la pesanteur,
\item Les valeurs numériques suivantes sont empruntées à une maquette de laboratoire:
\begin{itemize}
\item $S = 0.0154$m est la section des cuves,
\item $S_p = 5.10^{-5}$m est la section des tuyaux entre cuves,
\item $\mu_1 = \mu_3 = 0.5, \mu_2 = 0.675$  sont les coefficients de viscosité.
\end{itemize}
\end{itemize}
On régule, selon le paragraphe \ref{mfc}, grâce à deux systèmes ultralocaux \eqref{ultralocal} en parallèle, d'ordre $1$ et monovariables, de commandes et de sorties respectives $u_1$, $u_2$ et $y_1$, $y_2$. Pour les deux, $\alpha = 100$, $K_P = 0.5$. Un bruit additif, blanc, gaussien, centré, d'écart-type  $0,5.10^{-3}$, parasite les sorties. Les figures {\ref{C_u1}} à {\ref{C_l2}} rapportent les simulations obtenues avec un échantillonnage d'$1$s. Les poursuites de trajectoires sont excellentes (se rapporter à {\ref{C_e1}} et {\ref{C_e2}}). En dépit du couplage physique évident, une variation de consigne dans la cuve $1$ n'influence guère la poursuite dans la cuve $2$.

\begin{figure}[htbp]
\centerline{\includegraphics[width=5.19in,height=2.83in]{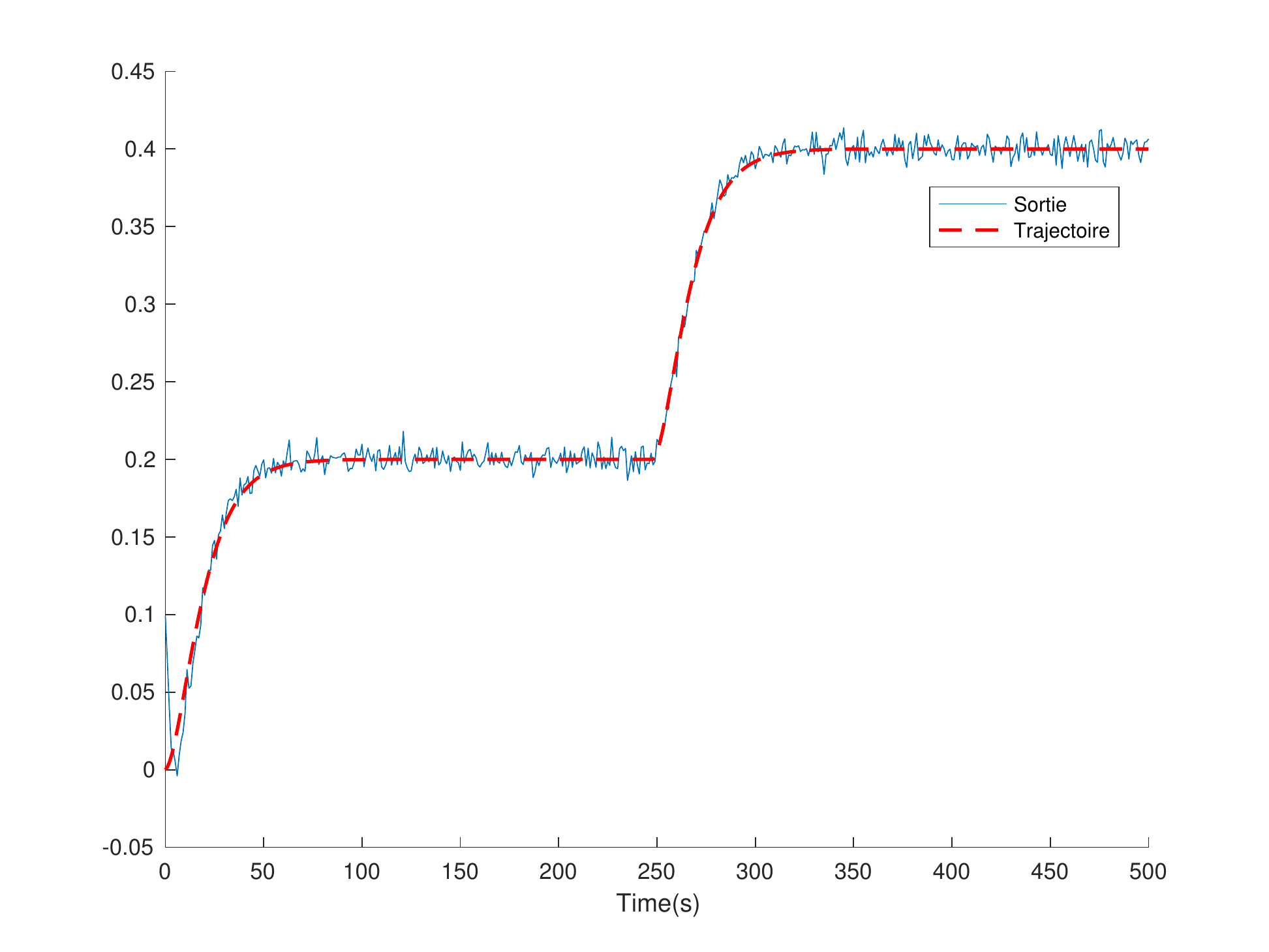}}
\caption{\textit{\sffamily Evolution du niveau dans la cuve 1}}
\label{C_l1}
\end{figure}

\begin{figure}[htbp]
\centerline{\includegraphics[width=5.19in,height=2.83in]{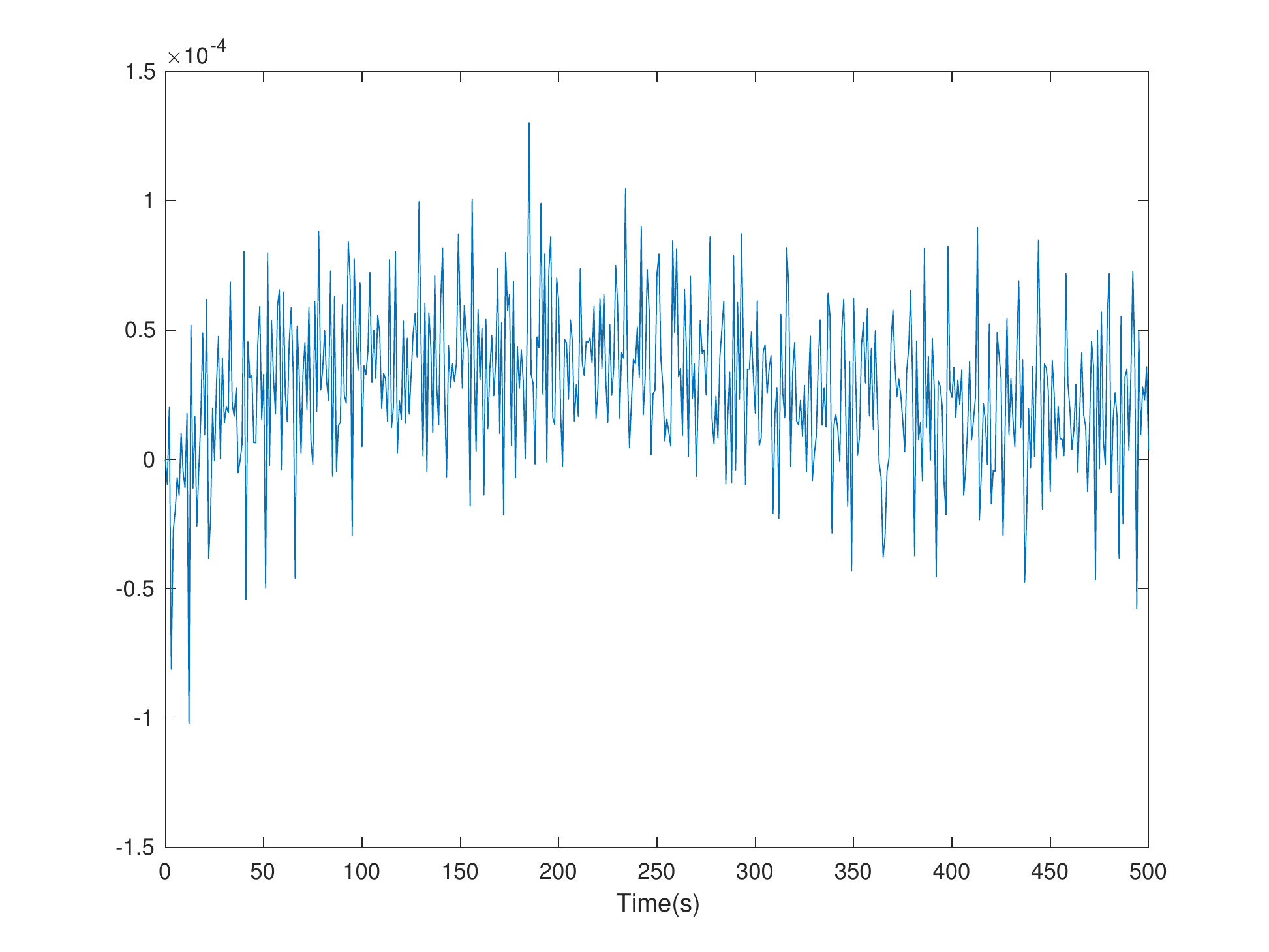}}
\caption{\textit{\sffamily Commande de la pompe 2}}
\label{C_u2}
\end{figure}

\subsection{Paramètres répartis: une équation aux dérivées partielles}
Soit l'équation de la chaleur en une seule dimension d'espace
\begin{equation}\label{chaleur}
\frac{\partial w}{\partial t}=\frac{\partial^2 w}{\partial x^2}
\end{equation}
où
\begin{itemize}
\item $0\leqslant x \leqslant 1$;
\item  $w(t,1)=u(t)$ est la commande;
\item $w(t,0)=0.5$ est la condition à l'autre bord;
\item $w(0,x)= 0.5+(u(0)-0.5)x$ est la condition initiale.
\end{itemize}
On désire ma\^{\i}triser l'évolution de $w(t,x)$ en $x = 1/3$ (voir figure {\ref{EDPy}}). Ici, $\nu = 1$, $\alpha = K_P = 10$. Réalisées avec un bruit additif, blanc, gaussien, centré, d'écart type $0.01$, et un échantillonnage de taille $T_e=10$ms,
les simulations des figures {{\ref{EDPu}}, {\ref{EDPy}}, {\ref{EDPe}} et {\ref{EDP3d}} présentent d'excellents résultats.

\begin{figure}[htbp]
\centerline{\includegraphics[width=5.19in,height=2.83in]{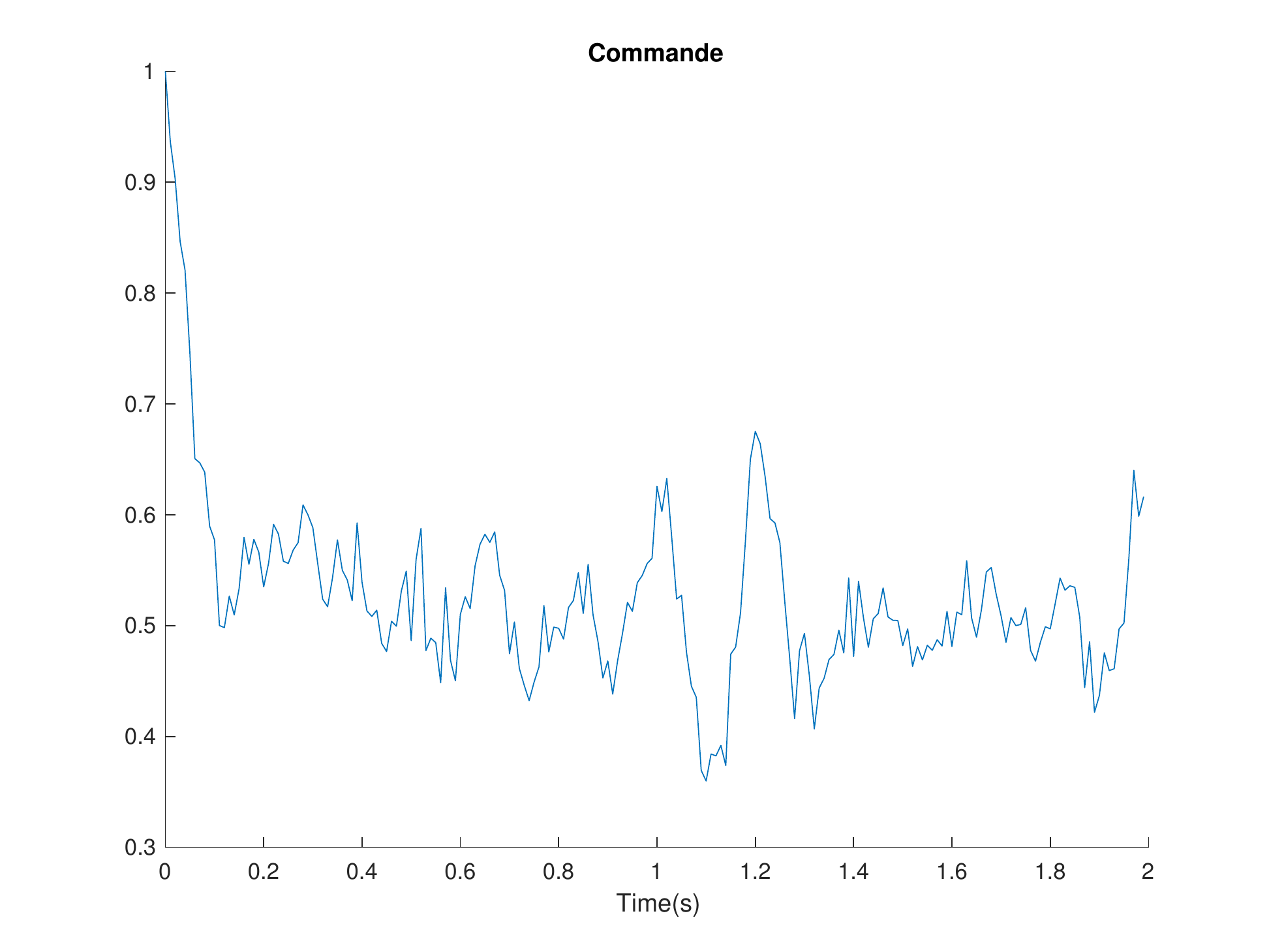}}
\caption{\textit{\sffamily Commande }}
\label{EDPu}
\end{figure}
\begin{figure}[htbp]
\centerline{\includegraphics[width=5.19in,height=2.83in]{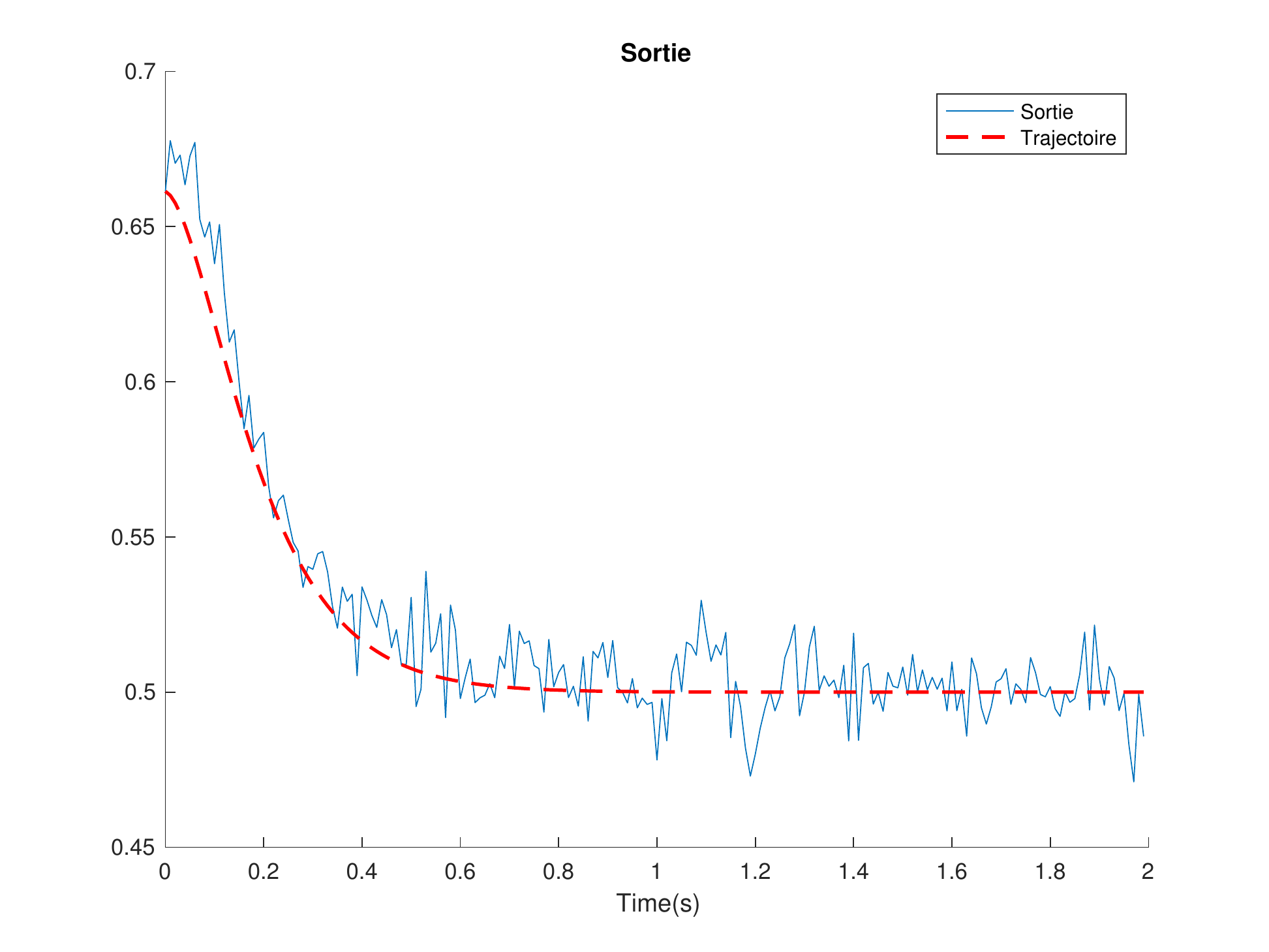}}
\caption{\textit{\sffamily Sortie }}
\label{EDPy}
\end{figure}
\begin{figure}[htbp]
\centerline{\includegraphics[width=5.19in,height=2.83in]{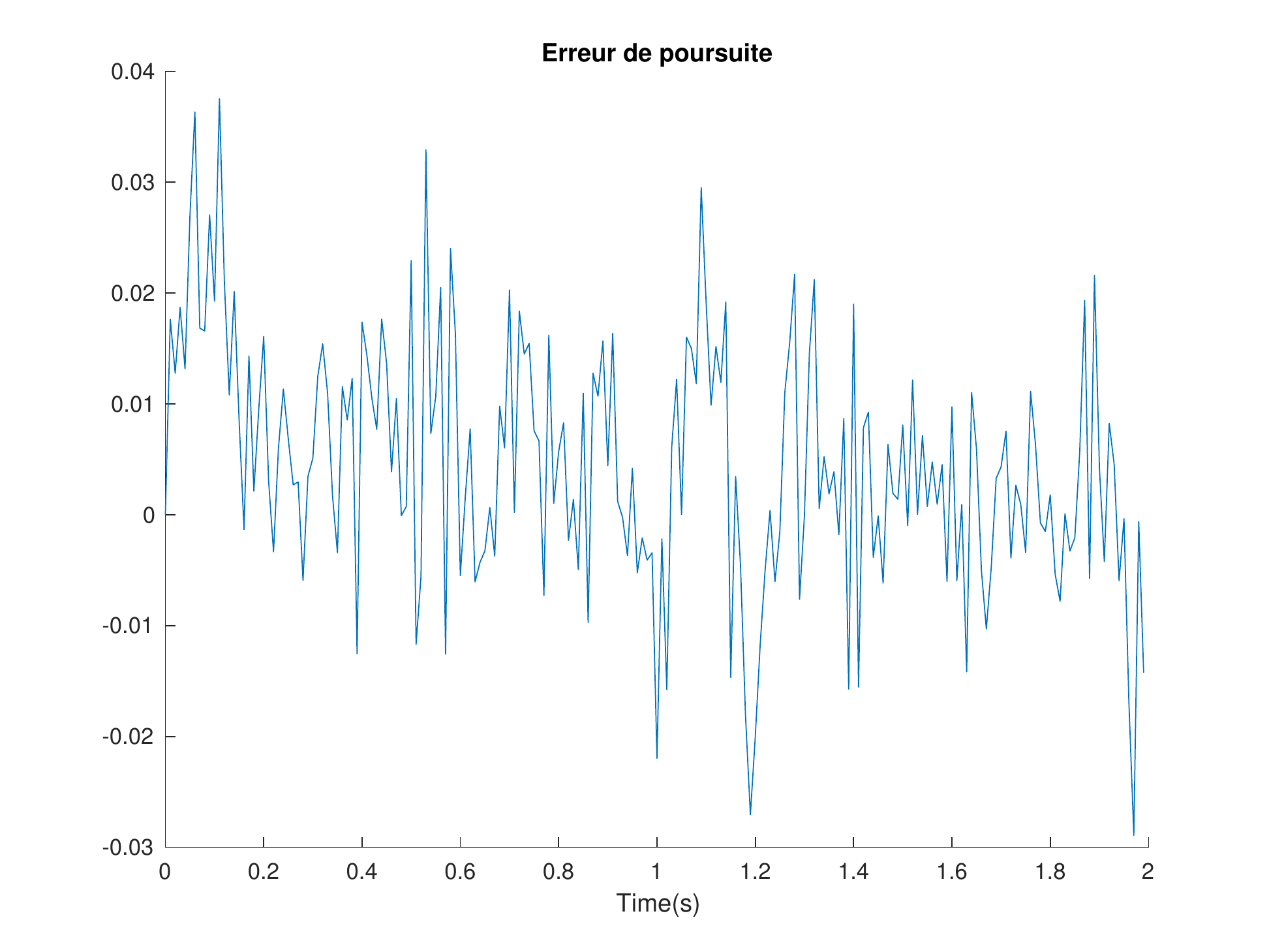}}
\caption{\textit{\sffamily Erreur de poursuite }}
\label{EDPe}
\end{figure}

\begin{figure}[htbp]
\centerline{\includegraphics[width=5.19in,height=2.83in]{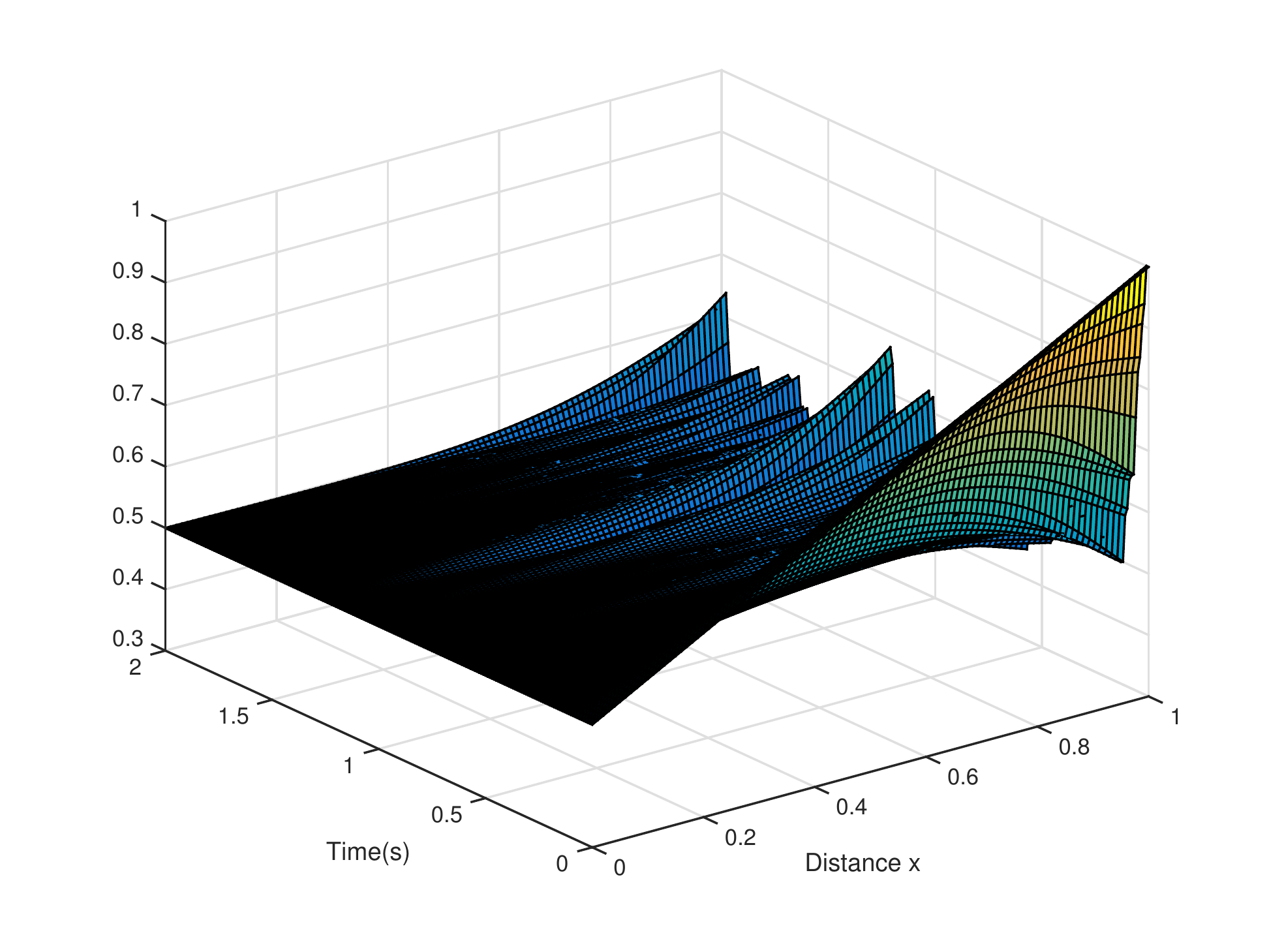}}
\caption{\textit{\sffamily Evolution de la température de poutre }}\label{EDP3d}
\end{figure}

\begin{remark}
En calcul opérationnel (voir, par exemple, \cite{erde}), \eqref{chaleur} conduit à considérer des exponentielles de la forme $\exp (\pm x \sqrt{s})$. Nul besoin avec MFC de développements originaux pour la dérivation fractionnaire $s^{1/2}$ et, on en fait la conjecture, pour toute dérivation non entière rencontrée en pratique. Ce n'est pas le cas avec l'ADRC (voir, par exemple, \cite{li}). 
\end{remark}

\section{Conclusion}
Le MFC a déjà permis des avancées conceptuelles d'une importance certaine:
\begin{enumerate}
\item On fournit une explication, en \cite{andr}, \cite{ijc13}, à l'universalité des PID et à leurs faiblesses, et ce pour la première fois.
\item En dépit de performances parfois insatisfaisantes, ALINEA\footnote{Acronyme d'\og{\underline{A}sservissement} {\underline{LIN}\'{e}aire} {d'\underline{E}ntr\'{e}e \underline{A}utorouti\`{e}re}\fg.} est l'algorithme le plus usité de régulation des accès d'autoroutes. Aucune justification n'en avait été publiée jusqu'à \cite{alinea}.
\item Réponse \cite{mathmod} à une question \cite{karin} sur la \og{compensation dynamique}\fg, en \og{biologie systémique}\fg, ou \emph{systems biology}.
\end{enumerate}
Aucun succès de cet ordre ne peut, à notre connaissance, être porté au crédit de l'ADRC. 

Depuis \cite{ziegler1, ziegler2}, des retards sont introduits pour faciliter le réglage des gains des PID. Ils perdent tout sens  \cite{ijc13} avec les correcteurs intelligents associés au MFC. Avec, par contre, des retards physiques \og{assez importants}\fg, aucune solution définitive ne semble s'imposer à l'heure où ces lignes sont écrites, en dépit d'avancées prometteuses, tant en MFC (voir \cite{doublet1}, \\ \cite{doublet2}, où $\alpha$ en \eqref{ultralocal} devient variable, et \cite{thabet}, \\ \cite{delayzhang}) qu'en ADRC (voir, par exemple, \cite{guo}, \\ \cite{sira3}, \cite{xia}, \cite{zhao}). Dominer cette question est un point clé. 

Quant aux systèmes \textit{a priori} modélisés par des équations aux dérivées partielles, la situation reste assez floue, même si MFC compte déjà quelques réussites concrètes incontestables (voir, par exemple, \cite{edf}\footnote{Il s'agit d'une installation hydroélectrique, souvent modélisée, comme les canaux, par une équation de Saint-Venant. La solution proposée, brevetée par EDF (\'Electricité de France) et l'\'Ecole polytechnique, a reçu le prix de l'innovation 2010, catégorie \og{brevet}\fg, décerné par l'\'Ecole polytechnique.}).

\end{document}